# White Paper on 6G Drivers and the UN SDGs

**30 April 2020**

## Executive summary

The commercial launch of 6G communications systems and United Nations' Sustainable Development Goals (UN SDGs) are both targeted for 2030. 6G communications is expected to boost global growth and productivity, create new business models and transform many aspects of society. The UN SDGs are a way of framing opportunities and challenges of a desirable future world and cover topics as broad as ending poverty, gender equality, climate change and smart cities. The relationship between these potentially mutually reinforcing forces is currently under-defined. Building on the vision for 6G, a review of megatrends, on-going activities on the relation of mobile communications to the UN SDGs and existing indicators, a novel linkage between 6G and the UN SDGs is proposed via indicators. The white paper has also launched the work of deriving new 6G related indicators to guide the research of 6G systems. The novel linkage is built on the envisaged three-fold role of 6G as 1) a provider of services to help steer and support communities and countries towards reaching the UN SDGs', 2) an enabler of measuring tool for data collection to help reporting of indicators with hyperlocal granularity, and 3) a reinforcer of new ecosystems based on 6G technology enablers and 6G network of networks to be developed in line with the UN SDGs that incorporates future mobile communication technologies available in 2030. Related challenges are also identified. An action plan is presented along with prioritized focus areas within the mobile communication sector technology and industry evolution to best support the achievement of the UN SDGs.

# Contents

# List of contributors

**Editor in Chief:**

- Marja Matinmikko-Blue, University of Oulu, Finland (email: marja.matinmikko at oulu.fi)

**Chapter Editors:**

- Sirpa Aalto, University of Oulu, Finland (Chapter 1)
- Muhammad Imran Asghar, Aalto University, Finland (Chapters 1 and 5)
- Hendrik Berndt, WWRF, Germany (Chapters 2 and 6)
- Yan Chen, Huawei Technologies, Canada (Chapters 2 and 4)
- Sudhir Dixit, Basic Internet Foundation, USA (Chapter 6)
- Risto Jurva, University of Oulu, Finland (Chapter 5)
- Pasi Karppinen, University of Oulu, Finland (Chapter 3)
- Markku Kekkonen, University of Oulu, Finland (Chapter 2)
- Marianne Kinnula, University of Oulu, Finland (Chapter 3)
- Panagiotis Kostakos, University of Oulu, Finland (Chapter 4)
- Johanna Lindberg, Luleå University of Technology, Sweden (Chapter 4)
- Edward Mutafungwa, Aalto University, Finland (Chapter 5)
- Kirsi Ojutkangas, University of Oulu, Finland (Chapter 4)
- Elina Rossi, University of Oulu, Finland (Chapter 4)
- Seppo Yrjölä, Nokia & University of Oulu, Finland (Chapter 3)
- Anssi Öörni, Åbo Akademi, Finland (Chapters 2, 5 and 6)

**Contributors:**

- Petri Ahokangas, Oulu Business School, Finland
- Muhammad-Zeeshan Asghar, University of Jyväskylä, Finland
- Fan Chen, ZTE Corporation, China
- Netta Iivari, University of Oulu, Finland
- Marcos Katz, University of Oulu, Finland
- Atte Kinnula, VTT Technical Research Centre of Finland Ltd., Finland
- Josef Noll, University of Oslo & Basic Internet Foundation, Norway
- Harri Oinas-Kukkonen, University of Oulu, Finland
- Ian Oppermann, NSW Government, Australia
- Ella Peltonen, University of Oulu, Finland
- Hanna Saarela, University of Oulu, Finland
- Harri Saarnisaari, University of Oulu, Finland
- Anna Suorsa, University of Oulu, Finland
- Gustav Wikström, Ericsson Research, Sweden
- Volker Ziegler, Nokia, Germany

# 1. Introduction

The United Nation's Sustainable Development Goals (UN SDGs) were introduced in 2015. There are 17 goals included in the Agenda 2030. The agenda itself together with the targets and the indicators forms the action plan for achieving the goals. The goals are framed to address global challenges including climate change, poverty and inequalities [UN 2015]. The UN SDGs also provide a high-level guide as to how governments and companies should plan their future strategies when striving for sustainable ways to live, produce and consume.

This 6G White Paper looks at how 6G research and development are connected to the UN SDGs. In the Vision-chapter, the operational environment of 6G is defined. The starting point is that 6G will open a new era of 'Internet of Intelligence' with connected people, things and intelligence. The UN SDGs are a response to global megatrends which will also affect 6G research and development. In the Megatrends-chapter, the trends for sustainable development of 6G are identified and the effects of these trends are examined. The fourth chapter begins with an outline of prior work which has mapped the linkage between the UN SDGs and ICT or mobile communications industries. The chapter presents a renewed linkage between 6G and the UN SDGs, as well as identifying relevant indicators. Chapter five identifies technical, regulatory and business-related challenges and obstacles. An action plan is presented to guide the development of 6G and support the linkage between 6G and UN SDGs. Finally, concluding remarks present a set of research questions.

In the first 6G White Paper from 2019, all the UN SDGs were identified as important targets for future 6G networks [6G Flagship 2019, p. 5]. The White Paper also defined societal and business drivers for 6G using a PESTLE analysis framework (Political, Economic, Social, Technological, Legal and Environmental). In the current White Paper, the connection between 6G and UN SDGs is further elaborated by investigating the existing linkages between the mobile communications and UN SDGs. 6G brings new opportunities help tackle global challenges, but at the same time all stakeholders should take into account the ethical premises of the UN SDGs.

According to the UN Global Sustainable Development Report 2019 [UN 2019b, pp. xxiii-xxxii], a number of cross-cutting factors, which are relevant to all SDGs, are critical to successfully achieving the "Agenda 2030":

- Human well-being and capabilities obstacles
- Sustainable and just economies
- Food systems and nutrition patterns
- Energy decarbonization with universal access
- Urban and peri-urban development
- Global environmental commons

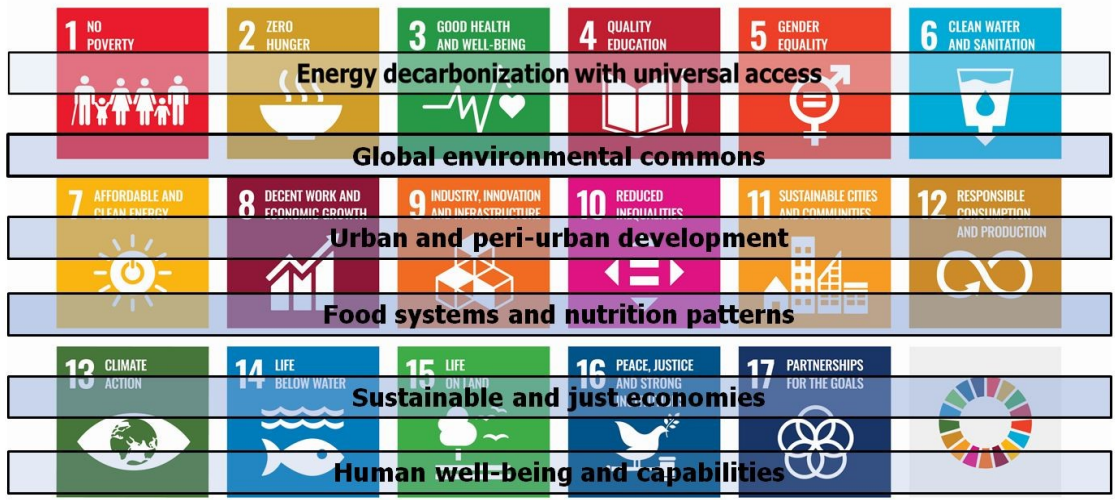

**Figure 1-1.** The UN SDGs and six cross-cutting factors ("entry-points") mentioned in the UN Global Sustainable Development Report 2019 [UN 2019b]. The entry points are thus factors which concern all the SDGs.

The rationale for highlighting these critical success factors (or "the entry points") is that progress on all the UN SDGs "will only be achieved if important trade-offs are addressed and transformed" [UN 2019b, p. 27]. The six critical cross-cutting factors should be considered when addressing any of the goals, because it is argued that the goals can only be achieved by a comprehensive approach (See Fig. 1-1). It is important to keep in mind that the advancement of one goal may advance other goals but may also have adverse effects on others. When considering 6G and mobile communications technology and its development, it is important to keep in mind that the developed technology itself should be sustainable and used in a sustainable way. The mobile communications community must pay attention to matters such as overall energy consumption, use of non-toxic materials, environmentally friendly base stations (including locating them where they do not adversely impact the environment), and sustainable supply chains. While not all adverse interactions between goals can be avoided, the trade-offs that are made should be openly addressed and mitigated . This is where the entry points offer useful tools so that adverse interactions between the goals are minimized. Besides technology, 6G has a clear potential to directly contribute to the UN SDGs by addressing digital inclusion and societal empowerment. For example, placing a base station in a remote area may advance the quality of life and work opportunities in peri-urban or remote locations, but have negative effects on the surrounding wildlife. The entry points offer a more a holistic approach to avoid such adverse interactions between the goals and have proved to be useful when examining how 6G can help achieve the UN SDGs (see Chapter 4).

The so called "levers" of: governance; economy and finance; individual and collective action; and science and technology; are highlighted in the Global Sustainable Development Report [UN 2019b, p. 29-36]. These levers impact the UN SDGs through these recognized entry points. Science and Technology, which is a pivotal lever from the perspective of 6G development, can bring about transformations in the entry points. Disruptive technologies can help overcome critical gaps in the delivery of cross-cutting initiatives that stem from the synergies of the UN SDGs.

The mobile communication sector plays an important role in societies around the world, and its linkage to the UN SDGs is many-fold. Mobile communications can significantly contribute to the achievement of the SDGs by offering infrastructure and access to digital services that will result in growth, efficiency and sustainability [ITU; GSMA 2018]. This is particularly true for economies where existing services are limited or existing service infrastructure is poor. Digitalization of services delivered through mobile communications networks has shown real benefit in developing economies in particular driving the uptake of micro-banking and micro-finance, micro energy grids and market creation [McKinsey, 2016]. Advancement of technologies for 6G may contribute further to the aim of reducing environmental impact through the 6G vision of reducing energy consumption of operating the networks including sue of zero energy devices with impact across industries and sectors.

The core principle of the Agenda 2030 is to "leave no one behind", and mobile technologies are the core for connectivity and Internet access [UN 2019a, p. 23]. The approach to achieving the goals will different around the globe because societies themselves are very different though 6G has the potential to be the global enabler. Universal access to information is now recognized as a crucial factor when addressing the ways to achieve the UN SDGs. The role of mobile communication in digital empowerment is very powerfully articulated by UN General Secretary António Guterres when he inaugurated the High Level Panel on Digital Cooperation[1]. The Panel's report, presented in June 2019, addressed a set of recommendations regarding, for instance, affordable access to digital networks[2], as well as digital and financial inclusion.

The UN SDGs also require consideration of external costs and bringing together a wide range of stakeholders to work towards the common goals. As of today, there are still major technological bottlenecks and constraints in achieving the UN SDGs' targets including: Opening up, democratizing and improving data; breaking down information silos; shifting the existing paradigm of policy making, which is largely based on intuition; towards an evidence driven approach enabled by big data; leveraging advanced tooling and prediction capability of machine learning and artificial intelligence while assuring privacy, trust and security of the users [European Commission, COM 2020, 65; EBA report 2020].

This White Paper aims to establish a linkage between 6G and the UN SDGs. In doing so, the scope of considered is not restricted to cellular mobile communication networks but the impact of communications networks more widely. The vision for 6G holds special promise as it includes the powerful concept of the network acting as a sensor, creating a new paradigm of information collection and sharing. Access to affordable, high-quality, broadband Internet is a fundamental, common denominator factor for achieving the Agenda 2030 goals.

---

[1] High-Level Panel on Digital Cooperation, https://www.un.org/en/digital-cooperation-panel/

[2] Notably recommendation 1A: "We recommend that by 2030, every adult should have affordable access to digital networks, as well as digitally-enabled financial and health services, as a means to make a substantial contribution to achieving the SDGs…" and recommendation 1B: "We recommend that a broad, multi-stakeholder alliance, involving the UN, create a platform for sharing digital public goods…".

## 2. Our 6G vision

Our vision of 6G accentuates a development that will provide much more than innovative use case driven mobile communication solutions but rather serve higher societal ambitions. Foremost amongst them are the UN SDGs. We foresee that our future society will be increasingly digitized, hyper-connected and globally data driven as outlined in [6G Flagship 2019]. Utilizing those opportunities, our vision allows for 6G to contribute to accomplishments such as helping people, society and economy adapt to the digital age -emphasizing of connecting the unconnected - strengthening the economy in a way that works for people and enhances the people's well-being, and supporting our planet through "Green Deals" and extreme energy efficiency. Policy frameworks and regulations have to be aligned, work hand in hand and have to reach up to the task level required to create the 6G future. Our societal ambitions are huge, including "societal readiness" on all levels for the digital future. It incorporates help for disaster reliefs such as pandemics and similar threats to be minimized through contributions from 6G development.

Fig. 2-1 illustrates the relationship between the UN SDGs and our 6G vision. 6G aims at providing ubiquitously several capabilities or services, as seeing on the left side of Fig. 2-1. These include *connectivity*, *sensing*, *actuating*, *intelligence* and *processing* capabilities/services. These will be essential building blocks to support the UN SDGs. However, in our 6G Vision some building blocks exist already or are part of the 5G evolutionary journey to 2030. Possible examples are context awareness and situational reasoning. It should be noted that all those focus areas embrace not only technical characteristics, but their potential should be understood in the wider context of transformational services, new ecosystems and value chains.

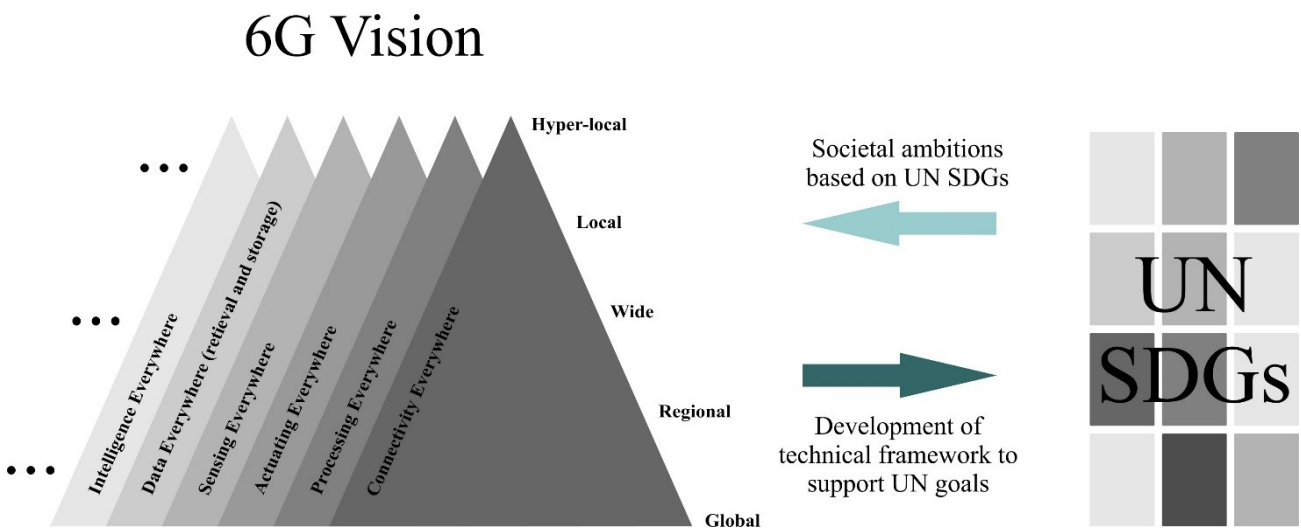

**Figure 2-1.** Relationship between the UN SDGs and the elements of 6G vision.

Fig. 2-2 summarizes the aforementioned connections between 6G and the UN SDGs. By supporting the UN SDGs, 6G aims at taking a leading role by a) *empowering the people* by providing services and solutions to individuals and society, b) *sensing the environment* by extensive hyper-local measuring associated to essential indicators and c) *strengthening the world* by reinforcing the ecosystem according to the SDGs. These three roles of 6G are further elaborated in Chapter 4.

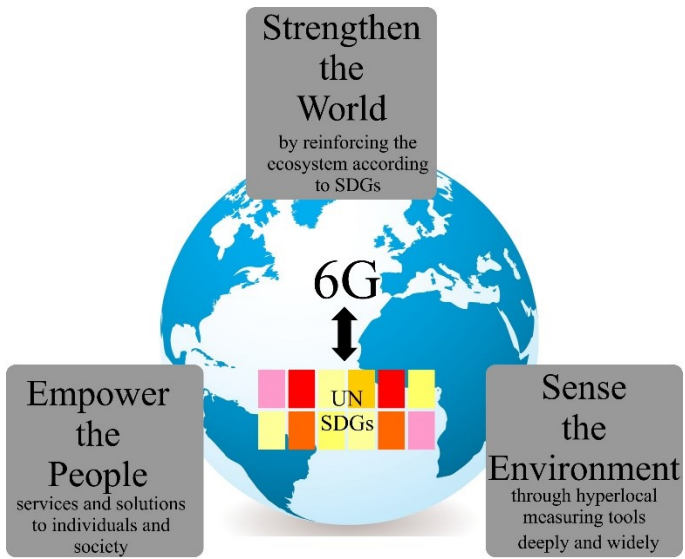

**Figure 2-2.** Three pillars linking 6G with the UN SDGs.

Exemplified by the current COVID-19 pandemic, prioritizing use cases that have the greatest potential to improve the lives of people around the globe could lead into rapid solutions if different stakeholders are prepared to work in unison. During the 2020 pandemic, we have seen automotive companies developing emergency ventilators for critically ill patients in intensive care. Conversely, the inability to collaborate could lead to chaos, which also seems to be present at some level during the current crisis. The choices people make in their everyday lives - whether common citizens, politicians or CEOs of global companies - do matter. We as humans often forget that open collaboration is essential, even though we may disagree on views. Therefore, we include in our vision that 6G development should facilitate open collaboration and open standardization between different stakeholders in multi-disciplinary manner, ultimately to create true partnerships for the benefits of all.

The way in which data is collected, processed, transmitted and consumed within the wireless networks will be a key driver for 6G as highlighted in [6G Flagship 2019]. Exploiting these data, ensuring privacy at the highest standard, we envision 6G will open a new era of an 'Internet of Intelligence' with connected people, connected things, and connected intelligence. Artificial intelligence (AI) will become a native feature of 6G. The AI reasoning and inferring capabilities will be embedded everywhere in the 6G networks to drastically enhance the network capability, and the communication architecture will be revisited to facilitate the collection and spread of intelligence.

Many widely anticipated future services would be critically dependent on instant, virtually unlimited wireless connectivity. Resources, specifically through utilization of higher and broader spectrum and larger arrays of antennas will continue to be exploited. 6G will provide a framework of services, including communication service where all user-specific computation and intelligence may move to the edge cloud. Natively integrating sensing with communication using specialized virtualization techniques will bring another leap on the network capabilities. Capacity, latency reduction, and positioning accuracy will improve, providing truly immersive aural, visual, haptic and sensory communication experience. The support of energy harvesting mobile devices involving charging via radio waves or laser beams and through utilizing energy from solar cells or body movements will also play a major role.

To ensure coverage including all remaining unconnected areas in the world, 6G rely on seamlessly integrating non-terrestrial networks such as UAVs (unmanned aerial vehicles), drones, and very low orbit satellites into the cellular communication systems. New hardware, new devices, and new user interfaces will come into play, exploiting all possible types of renewable energy, and transforming the way human beings interact with the digital world.

While pushing many technological boundaries in the journey to 6G, the UN SDGs are an important lens to help prioritize development. Technology standards which support those use cases that hold the highest promise for improving human lives and protecting the environment need to be advanced first. This journey will need to consider mechanisms to help policymakers around the globe to advance the UN SDGs by positively influencing the behavior of people. The 6G ecosystem will provide solutions that empower individuals and communities to adopt self-correcting processes and steer the actions towards long term sustainability. 6G will also provide the means for sufficient data collection - with strict privacy protection rules and in line with values of our civil society, corporate accountability and government transparency - to enable the monitoring of SDG indicators that are or will become relevant in the near future.

Fig. 2-3 depicts the three high-level aspects in the 6G vision, namely a) people/society, b) machines and things and c) resources. These are in line with the 6G goal *everyone connected*, *anything connected*, and *any resource*, respectively. Resources refer to assets that can be exploited either directly or in a shared fashion. These resources can be tangible or intangible. Tangible resources include concrete physical assets, whereas intangible resources consider immaterial resources such as knowledge and energy. 6G will connect these aspects using both local and wide wireless networks to create a hyperconnected world bringing communities every closer to a globally connected world.

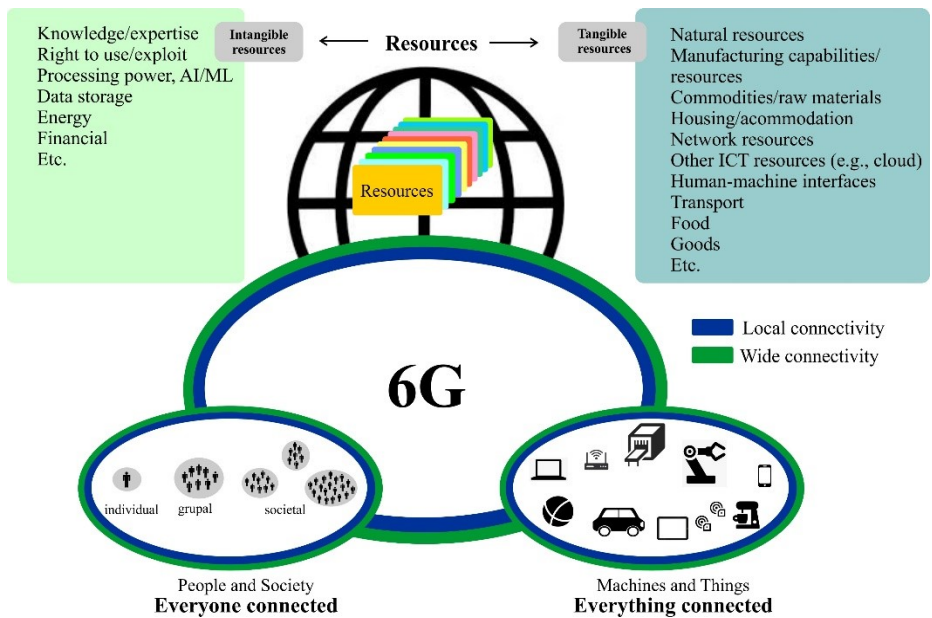

**Figure 2-3.** Key high-level 6G aspects interconnected by wireless and mobile networks.

Since the world is not homogeneous, any development to achieve the 6G vision must support an open architecture, crowd sourcing, and enablement of third-party application and service eco-system to meet local needs. At the same time 6G will not provide unlimited capabilities but rather provide for the real needs, such as high-capacity networked islands where and when needed and provide a sustainability perspective.

The operational environment for international 6G research and development consists of multiple stakeholders with different goals and roles including, e.g. research and educational organizations; governmental, regulatory and standardization organizations; users; industry; and verticals, as depicted in Fig. 2-4. Research and educational organizations have an important role in driving the 6G research and educating professionals, see e.g. [6G Flagship 2019]. Governmental organizations including national level decision makers create the incentives and national conditions for 6G networks and services. Regulatory bodies at various levels are in charge of many critical 6G deployment aspects including regional and international spectrum matters. Standardization organizations enhance the adoption of necessary technology for sustainable development, see [ISO 2018; IEC 2020]. Industry encompasses various stakeholders such as mobile network operators (MNOs) and digital service providers, all investing in the infrastructure, and other telecommunication industry that develops the 6G technology and applications. Verticals include stakeholders from e.g. automotive, health, energy and other sectors that use the technologies in specific application areas. Given the challenges and economic impact of adopting the UN SDGs, various verticals need to engage earlier in the process compared to development of previous generations of mobile communications. Also special purpose 6G networks will be operated in verticals and the various applications tailored for verticals' use will be running on the 6G networks. Finally, we want to remind about the role of the users of the future systems. To prevent digital divide, people need to have the devices required to use the services and they also need to have skills to use those as well as the available services. Furthermore, people need to understand technologies and they need to learn appropriate technological skills to actively transform disruptive technologies into advancing empowerment.

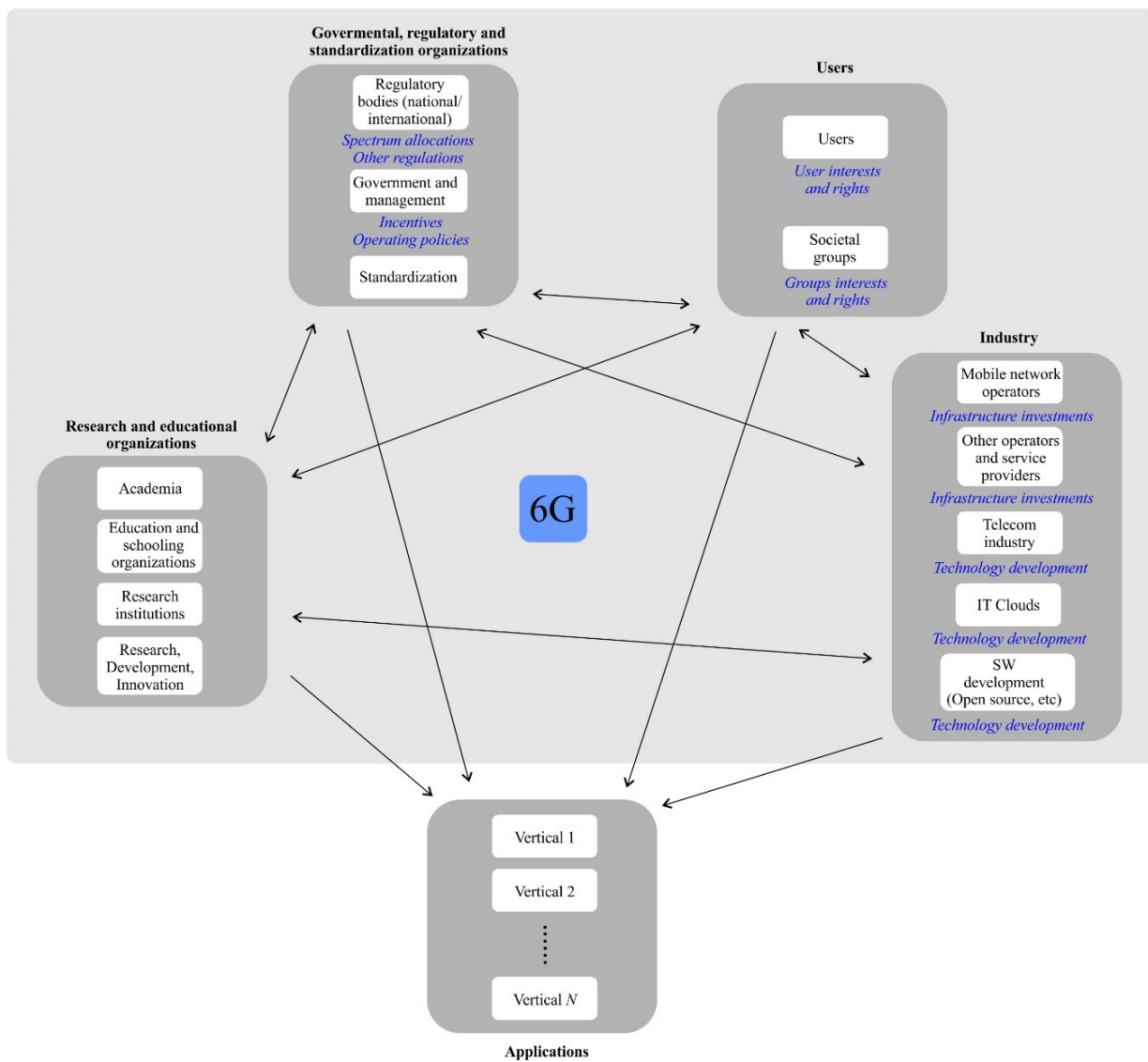

**Figure 2-4**. Operational environment for 6G.

# 3. Megatrends driving the 6G research

6G research and development are unfolding against the background of global megatrends that will shape our world over the coming decades. Megatrends are change-related phenomena and transformative, global forces that define the future world. They have impact on businesses, societies, economies, cultures, and personal lives. Though the COVID-19 pandemic dominates the news, climate change is probably the most topical today, but phenomena including population growth and demographics, increasing environmental pollution or global competition for resources are also current megatrends [Dufva 2020]. For example, the United Nations Office at Geneva (UNOG), as well as the Government of Norway, have pointed out the challenge of Digital Divide as one of the global challenges. The governmental report [GovNorway 2020] on "Digital transformation and development policy" identified the following four barriers to digitalization (i) access, (ii) regulation, (iii) digital competence and (iv) exclusion, as priorities for the government. In a similar matter, United States Agency for International Development (USAID) has added "last mile connectivity" as the main priority for development policy. From the perspective of digitally empowering the users globally, it should be emphasized that digital inclusion is the catalyst for achieving the UN SDGs, implying that we also need to address how everyone in the society can benefit from basic information in the digital world available free on the web. Furthermore, certain UN SDGs require global cooperation to be successful leaving aside national interests, such as environment and ecological reconstruction, controlling spread of diseases.

## 3.1 Quintuple Helix model for sustainable development of 6G

The trends identified in the White Paper on Business of 6G in [6G White paper Business of 6G, 2020] were taken as the baseline and categorized according to the Quintuple Helix model [Carayannis et al. 2012]. The Quintuple Helix aims to support a win-win situation between ecology, knowledge and innovation, creating synergies between economy, society, and democracy. It can be seen as "a cooperation system of knowledge, know-how, and innovation for more sustainable development" [Carayannis et al. 2012].

In the Quintuple Helix Model (see Fig. 3-1), there are five **subsystems** (helices): Political system, Education system, Economic system, Natural environment, and Media-based and Culture-based public (civil society). Each subsystem has its own asset, **capital**, at its disposal and forming within the helix, resulting from the actions in the helix itself. **Knowledge** is in this model treated as a resource that circulates between the subsystems and changes to innovation and know-how in a society and for the economy. All subsystems are seen to influence each other through knowledge. Knowledge acts as an **input** to the subsystems, a resource for new **knowledge creation** in the subsystem. As an **output** of a subsystem, there is new knowledge in form of know-how that continues circulating the system, or new inventions. [Carayannis et al. 2012].

**Education system** contains all levels of education. Its asset is embodied in humans who either go through the system (students) or work within it (teachers, researchers etc.) as human capital, resulting from diffusion and research of knowledge. Knowledge creation within this subsystem is e.g. education in different educational institutes as well as research. **Economic system** consists of different elements that comprise the economic structure of a community, such as institutions, industries, companies, services, banks. Its asset is economic capital, i.e. everything that makes it possible to perform economically useful work. This includes for example machinery, production processes, resources, and money. **Natural environment** has as its asset natural capital, i.e. nature itself and the resources it provides, such as soils, water, air, living organisms, minerals, and metals. Assets **Media-based and culture-based public**, a combination of the civil society and media, are information capital and social capital that are integrated and combined in the subsystem. This capital includes on the one hand, information and media related assets such as television, newspaper, and information sharing networks, and on the other hand culture related assets such as e.g. art, traditions, values, and life-styles. The **political system**'s assets are political and legal capital. Political capital refers to the accumulation of resources and power to the politicians, parties, and other stakeholders, and legal capital is understood as the legal system and laws and regulations. [Carayannis et al. 2012]

In the quintuple helix model, circulation of knowledge between the subsystems is seen to happen within a nation-state but also between states [Carayannis et al. 2012]. Next, we go through megatrends we have identified as relevant from sustainable development of 6G point of view. The trends largely cross the boundaries of nation-states, thus the quintuple helix model needs to be understood in different levels as well, both at the nation-state level as well crossing these boundaries.

### 3.2 Identified megatrends

Following the Sitra's report of megatrends [Dufva 2020], there are five trends that are vital in the future: **1) the need for ecological reconstruction, 2) the strengthening of relational power, 3) the ageing and diversification of the population, 4) the redefinition of the economy**, and **5) technology is embedded in everything**. The megatrends in [Dufva 2020] have been derived to help in understanding of the visible changes in the world, especially from Finland's perspective, but are not limited to a single country. Next, we link these five megatrends to corresponding helices of quintuple helix model and add to that such trends identified in 6G white paper 8 [6G White paper Business of 6G, 2020] that fit this white paper. See Fig. 3-2 for the mappings and further details in the following sections.

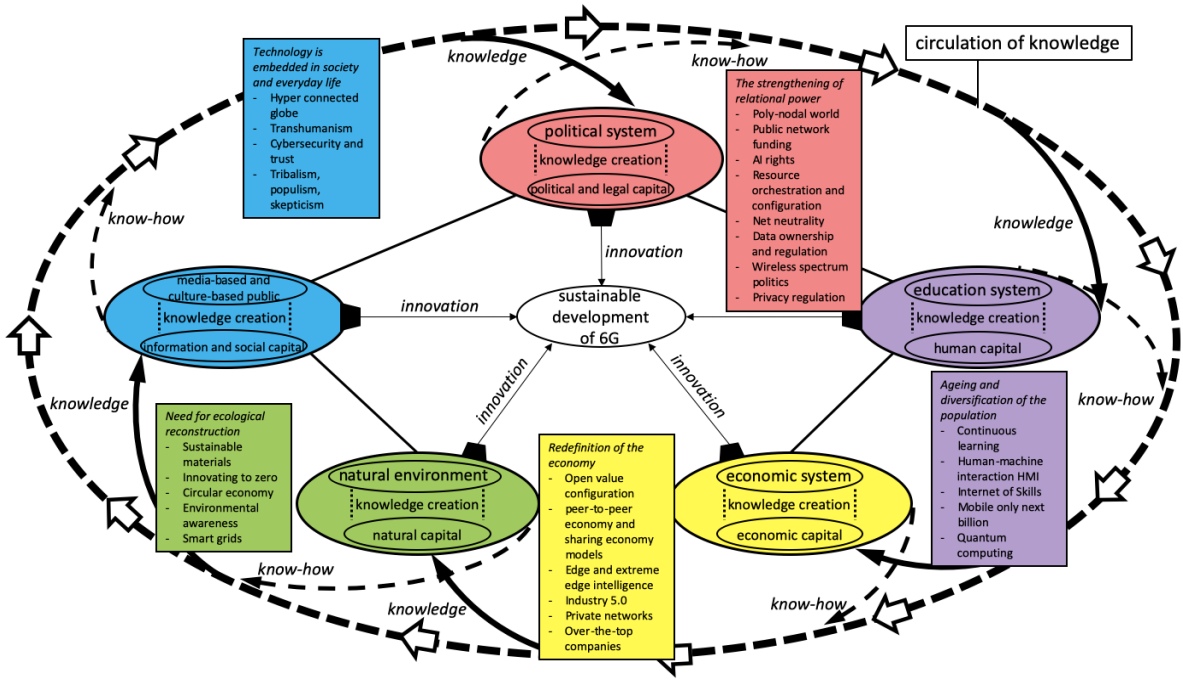

**Figure 3-1.** Megatrends in the context of quintuple helix model. Framework adapted from [Carayannis et al. 2012, p. 7].

### 3.2.1 Trends related to the natural environment

Arguably the key trend influencing our future is the need for **ecological reconstruction**: how do we respond to climate change, decreasing biodiversity, the dwindling availability of resources and waste-related problems? Other trends need to be viewed against this backdrop as well. For example, the political system should be able to make decisions quickly – but are they made through centralization of power or inclusive decision-making or political correctness?

Other trends in the natural environment system includes e.g. **sustainable materials,** which in turn contributes to **Innovating to zero** and **circular economy** trends. Towards 2030 companies will shift focus and develop products and technologies that innovate to zero, including zero-waste and zero-emission technologies. While technology offers **new solutions for the production of energy** it also simultaneously increases the demand for energy. This creates a conflict: to what extent does technology promote ecological reconstruction and to what extent does technology hinder it? 6G net positive impact and sustainability are expected to be through enabling increased efficiencies and improved environmental performance. Technologies of **computing will be miniaturized** to the extent that they sustain on the power generated through everyday human activity. The everyday activity of walking, jogging and everyday housework could produce energy to support the person's information devices which in turn would monitor person's vitals from time to time as well as cater to information and entertainment needs through over-the-top connectivity.

At the same time, **environmental awareness** among people and companies has increased, reflected in a growing number of people and communities changing their habits and companies taking corresponding actions to offer customer experience. Dissatisfaction with the current measures taken with respect to climate change and biodiversity has motivated a growing number of people to voice their opinions and participate in demonstrations. Responding to the increased environmental awareness requires changes in culture and practices and has been accompanied by a polarization of views.

In the future **smart grids** are extended to a variety of sectors including electricity, internet and healthcare. All of these are hyperconnected and completely automated. They serve as a middle layer between humans and natural environments enhancing the capabilities of both. These networks have been put together with public-private-personal ownership funding model with a view towards sustainable growth and usage of digital infrastructure.

### 3.2.2 Trends related to the political system

We are moving from multipolar world to a "poly-nodal" world, where **strengthening of relational power** is vital. In *geopolitics* the tension between the globalization, networking power and the urgency of the ecological reconstruction is linked to the balance between centralized decisions and the strengthening of inclusion and democracy. Towards 2030 *power configuration* is transforming from a *multi-polarized world to a poly-nodal* world where power will be determined in the economic, technological and cultural networks and interaction. Power is determined by relational influence and held not only by states but also by companies, regions, communities and transnational organizations. Societies are struggling to find a balance between **quick-moving decision making, community engagement and the reasserting of democratic values and commitments**.

*Public network funding* has traditionally been directed to unserved and underserved areas in terms of **broadband access and coverage**. Lately support for deployment programs has extended to policy areas, such as **smart city community development**, worksites and ecosystems (such as harbors and airports), advanced health services, logistics and transport, Smart Cities, public safety and critical infrastructure at length. Smart society builds dependable systems and communication where standardized data is utilized by walled garden platform monopolies across verticals. Smart city focus is extended to rural inclusion. 6G will transform urban and rural living existing at the intersection of geopolitics, growth of nationalism, rights to information transparency and information citizenry. *Resource orchestration and configuration* relates to the **power over development and adoption of innovations** and technology that ubiquitously embedded in society and our daily life. **Technology is** increasingly seen as a geopolitical issue of **power** and the questions of future resource orchestration power emerge: **who will own the continuously accumulating data**, who will get to **decide on technology** and who sets the **rules and regulation.**

*Data ownership and regulation.* **Access to data and data ownership are increasingly the major factors in value creation**, and **limiting such access is a means of control**. Creating a system that transforms how data is collected, shared and analyzed in real time can create strong drivers for future value, introduce novel stakeholder roles, but may also lead to serious privacy and ethical concerns over the location and use of data.

*Privacy regulation* is strongly linked with the rising trends of platform data economy, p2p sharing economy, intelligent assistants, connected living in smart cities, transhumanism and digital twins' reality at length. "**I own my data**" is expected to grow particularly in Europe based on **GDPR** evolution, though severe **differences in the global data privacy laws** are expected to be living on borrowed time. *Artificial intelligence rights* trends have **contrary interpretations**. Assuming the availability of appropriate data sets for training purposes, artificial intelligence has the ability to propose solutions to increasingly complex problems. AI can be the **source of great economic growth, shared prosperity, and the fulfilment of all human rights**. In the alternative future it drives inequality, inadvertently divides communities, and is even **actively used to deny human rights**.

*Wireless spectrum politics* and spectrum management in the 6G era will reveal a new level of complexity that stems from the variety of spectrum bands and spectrum access models with different levels of spectrum sharing. **Local deployments of networks** by a variety of stakeholders is expected to further grow in 6G. Furthermore, in national technology politics, **spectrum regulation** is used to gain competitive advantage. According to *Net neutrality* ruling **Internet access providers should treat all traffic equally** irrespective of sender, receiver, content, service, application or device in use. At the same time, the **5G** evolution is already developing a network that **can be extremely tailored** to a use case intending to treat traffic differently for each use case. This legislation creates uncertainties by **impacting companies' capabilities to create and capture value** in virtualized network-based services between telecom operators and cloud providers.

### 3.2.3 Trends related to the education system

**Diversification of the population**: We see two trends, both ageing in well-established societies, as well as young generations in developing economies trying to catch up to the welfare of established economies. In well-established societies, the population is not only ageing but also becoming more diversified in terms of backgrounds, opportunities and habits [Dufva 2020]. Highly functional educational system can help to provide equal opportunities for all. Continuous competence development is not required only from work force but all citizens. Learning capabilities will have a key role in the future, which creates pressure to re-evaluate educational structures and policies.

New technological innovations can revolutionize education in the future. Emerging opportunity of redefining **human machine and brain-UI interface** enables connecting people and the biological world in novel ways. Holopresence systems can project realistic, full-motion, real-time 3D digital twin images of distant people and objects into a room, along with real-time audio communication, with a level of realism rivaling physical presence. Images of remote people and surrounding objects are captured and transmitted over a 6G network and projected using laser beams in real time. The pervasive influence of artificial intelligence will not just reflect what something looks like but also its context, meaning and function, creating Internet of skills, Internet of senses and digital twins [6G Flagship 2019].

However, not all have equal opportunities to access latest educational technology. Trends in **mobile only next billion** means, how ubiquitous cheap phones and increasingly affordable network connections in mega cities and rural are helping another billion users join the internet and access applications and digital content increasing at non-English speaking markets[3]. For many, mobile is the primary or only channel for accessing the internet and services. With its unprecedented scale and growing impact on daily lives, mobile is a powerful tool also for achieving the SDGs and drive sustainable economic growth.

Academia is crucially important stakeholder not only when providing education but also fostering groundbreaking research. Alternative concepts and models for computing such as **quantum computing that** is at its best in sorting, finding prime numbers, simulating molecules, and optimization, and thus could disrupt segments like finance, intelligence, drug design and discovery, utilities, polymer design, AI and Big Data search, and digital manufacturing. For long the technology will be limited to selected industries, such as military, national laboratories, and aerospace agencies while other alternative approaches of compute to help handle greatly increasing level of parallelism in algorithms may be available more widely.

---
[3] www.basicinternet.org

### 3.2.4 Trends related to the economic system

**The redefinition of the economy**, is also linked to sustainable development: is the environment only to be regarded as a resource or should the economy aim to improve the state of the environment?

**Urbanization** will bring 5 billion people together to live in cities by 2030 occupying 3 per cent of the Earth's land but accounting for 80 per cent of energy consumption and 75 per cent of carbon emissions. 95 per cent of urban expansion in the next decades will take place in developing world where 883 million people live in slums today. Rapid urbanization is exerting pressure on fresh water supplies, sewage, the living environment, and public health. Cities are hungry, global economic engines and the economic powerhouses of the global economy. In 2015, 85% of global GDP was generated in cities. Cities are increasingly functioning as autonomous entities, setting social and economic standards. Urban identity will grow in importance compared to national identity.

**Open value configuration** and open source paradigm may provide a powerful avenue to reinvent civil society participatory process and regulatory capability. Utilizing **sharing** and **circular** economy trends **co-creation** utilize existing resources and processes to promote the stable interaction of mechanisms. As counterforces to the creation of platform monopolies decentralized platform co-operatives, the **peer-to-peer** economy and **sharing** economy models and the progress of a human-driven fair data economy has emerged. Towards 2030 **platform ecosystems** will provide an infrastructure on which innovation and **transaction platforms** are built. **Crowdsourcing** and crowdfunding are expanding the space for new forms of organization and innovation. Component prices have already decreased significantly, more sophisticated components will become cheaper, and new ones are invented. Low component prices provide possibilities for both local businesses as well as international ones. This encourages **frugal** innovations that are supported by the global do-it-yourself culture where sharing of blueprints and working processes is the norm to fight inequality in the world. In the heterogeneous society social networks and the trust and reciprocity they foster are highlighted from the perspective of well-being as well as working life. Novel **decentralized business models** do not necessitate a focal point but depicts the design of transaction content, structure, and governance to create value.

Via **edge and extreme edge intelligence** the proliferation of ever more powerful communication, computing and analytics resources at the edge of the network has turned the architecturally disaggregated 6G access networks into a rich service provisioning and access platform. Hyper-local services, such as augmented reality scenarios, do not require connectivity to distant service platform but, instead, perform better with local real time service access. Furthermore, the person will be supporting part of shared information processing and edge intelligence networks that address collective problems for humanity, such as genome sequencing (citizen science), through shared resources. The individual emerges as a node in the network of intelligence relations rooted in the local physical world while connected to the hyper-real 6G intelligence networks. This adopts a viewpoint of a public good through digital infrastructure of 6G supported through an ecology of information devices, products and services of IoT/IoE.

**Over-the-top companies** will utilize their customer data, cloud infrastructure and AI/ML capabilities to challenge traditional operator's customer relationship ownership as users perceive connectivity as basic utility. In addition to the media space OTT players are offering basic telco services such as voice or messaging and are active in growth areas, such as cloud space and services, competing with telcos for clients and revenue. They are **tying customers to their own ecosystems with carrier-neutral connectivity**, while making reliance on traditional operators a thing of the past.

Towards 2030 **industry 5.0** will allow collaborative *human-machine interaction (HMI)* across services and industries as human intelligence is in perfect harmony with advanced, cognitive computing. With real-time data, effective data monetization and digital automation of the manufacturing process, businesses will be able to shift focus towards generating higher revenues from servitization of products. **Advanced manufacturing capabilities** will help to overcome design complexities and with its ability to facilitate extreme long tail of mass customization and further return the control back to customers and in a haptic way. **Private networks** driven by industrial digital automation calls for standalone networks for high reliability, high performance in terms of both bandwidth and reliability, secure communications and data privacy, fulfilling business and mission critical needs.

### 3.2.5 Trends related to the media-based and culture-based public system

**The technology is an embedded enabler in** everything: in knowledge creation and circulation of knowledge. Technology is becoming a part of society and everyday life. The sense of community created by 6G technology and the ability to directly collaborate with others enables humans to participate and act in society in an unprecedented way. **Hyper connected** globe will continue to feel ever smaller in 2030: Globally 90% people will be able to read and have access to Internet and this trend is on the move. The aim is to recognize that 6G will transform urban and rural living existing at the intersection of geopolitics, growth of nationalism, rights to information transparency and information citizenry. Thus, once the infrastructure of 6G is in place, the content growth will be in terms of supporting multiple social and technological identities of people through a variety of media. This would require a mindful view on decision making and regulation of future, data, information, media and network usage in light of sustainable growth for economy. Thus, the human in 6G worlds will be increasingly sophisticated in terms of their media and service consumption while being rooted in their local economies. Connectivity is therefore not only virtual and digital, but also physical, such that the physical world will seamlessly meet the virtual world and, with novel ways of interacting with the human biological world. When 6G technology has penetrated most parts of the world, IoT devices and sensors controlled by AI are an integrated part of environment. Automatic collection of different kinds of data (from humans: functioning of human body, biometrics, biosensors, etc. as well as from our environment etc.) and its analysis are used for highly sophisticated products and systems that make people's lives easier and provide better user experience through convenience as everything is automated.

The need for **cybersecurity and trust** will be ubiquitous in the hyper connected world 2030. Even a temporary **loss of technology** may have, not only a productivity impact, but also a psychological impact on our lives. Furthermore, the **subversion or corruption of our technology could result in the disastrous harm to our lives and businesses** if, e.g., medical treatment devices deliver the wrong medication, educational systems teach propaganda, or home, on the autonomous move or work automation cause us injury or damage our products and businesses. In particular, expectations to protect and **safeguard society and critical infrastructures** from emergency situations by means of technological advancements are anticipated to grow.

Furthermore, with increasing polarization and the ageing and diversification of the population new tribes and communities will emerge around various imaginary groups representing **wide variety of values**, place of residence, political opinion, consumption choices or lifestyles. With weakened and fragmented future prospects, the **absence of togetherness and the polarizing effect of social media** have led to a rise in **populism, skepticism** towards changes in the environment and in the worst case, **extreme attitudes**. **Transhumanism** reflects the rise of technology-driven evolution at an unprecedented speed of change, propelling deeper questions into **what it is to be human** from biological, behavioral and human-machine evolutional perspectives. Cognitive intelligence revolution via ascendancy of sentient tools and further possibly human orchestrated self-directed selection in biological, neurological, and physical evolution.

# 4. Developing a linkage between UN SDGs and 6G and related indicators

The UN Agenda 2030 for Sustainable Development introduces 17 goals and 169 targets that are measured with 232 individual indicators as presented in [UN 2017]. Prior studies on the linkage between the UN SDGs and the mobile communications sector, or more generally the ICT sector, have identified that all the SDGs can be significantly contributed to. There are on-going activities related to mobile communications and UN SDGs carried out by governments, regulatory bodies, standardization organizations, trade associations, companies, etc. In the following we provide an overview of relevant on-going activities in different forums working towards the achievement of the UN SDGs, followed by our new linkage between 6G and the UN SDGs.

The global indicator framework [UN 2017] for the goals and targets of the 2030 Agenda for Sustainable Development defines 232 individual indicators where of seven are addressing ICT related topics. Overall, the indicators describe today's situation and very few are ICT specific, which calls for the stakeholders to look into the development of a new set of more specific indicators which does not exist currently and are applicable to future mobile communications systems. In particular, there is a need to develop new indicators to guide 6G research and development towards 2030.

## 4.1 Prior work on linking UN SDGs with mobile communications and indicators

International Telecommunication Union (ITU) as the UN specialized agency for ICT has identified the role of ICT in achieving the different UN SDGs, see [ITU]. Similarly, for several years the mobile communication industry itself has evaluated its impact on the achievement of the UN SDGs, see e.g. [GSMA 2018; GSMA 2019]. The prior work on the linkage between the UN SDGs and mobile communications and related indicators are discussed next.

### 4.1.1 Existing linkage between UN SDGs and mobile communications

The identified linkage between the UN SDGs and the mobile communications sector or more broadly the ICT sector are summarized in Table 4-1. Mobile industry's contribution to the achievement of the UN SDGs is primarily considered through three levels: 1) the deployment of infrastructure and networks forming the foundation for digital economy, 2) providing of access and connectivity allowing people to use mobile communications, and 3) by enabling life-enhancing services and relevant content for people [GSMA 2018]. Broadband Commission for Sustainable Development by ITU and UNESCO has set its own seven broadband targets for 2025 including making broadband policy universal, making broadband affordable, getting people online, achieving digital skill and literacy, using digital financial services, getting businesses online and achieving gender equality [Broadband Commission 2019]. Additionally, the ITU and Digital Impact Alliance (DIAL) have developed a framework for SDG digital investments to choose technological solution [ITU – DIAL 2019]. The framework consists of four interrelated layers: SDG targets defining high-level objectives, use cases defining the steps to achieve a business objective contributing to one or more SDG Targets, work flows defining generic business processes supporting the delivery of a use case, and ICT building blocks that enable work flows and use cases. The framework provides a method for mapping of ICT building blocks to specific UN SDGs.

Although the mobile communications community has recognized the important role that mobile communications play in reaching all UN SDGs, the approach taken so far has been mainly economically driven. A purely commercially-driven roll-out will not always serve the under-represented groups in the community, e.g. those who do not see the value in the Internet. Instead, community networks are often established as individuals see the necessity of connectivity, though the economical viability is not in place. Therefore, communities need to be made an integral part of any policy making and decision-making bodies.

In this white paper, the scientific community together with other actors further develops the linkages between the technological solutions and the UN SDGs, and maps out how mobile communications can enable the achievement of UN SDGs from a wider, societal perspective, paying attention to issues such as equality, both on the individual and institutional level. As an example, we may take education. On the individual level, it is vital to have access to schooling no matter where you live. On the institutional level, this means the society has to provide infrastructure in order for schooling can be arranged. This is especially the case if online access to classrooms is necessary, for example in remote areas, or in situations like the COVID-19 pandemic where students cannot physically attend school due to government restrictions. In practice, institutions need to offer access to online services, and students need a mobile device. Technological solutions and services, and the technology needed, is developed and provided by the academic community and the industry. Communications technology industry, societies and public institutions have to join forces to achieve the goals the UN has set for education.

**Table 4-1.** Existing linkage between the UN SDGs and mobile communications/ICT.

| The UN SDGs | Existing linkage with mobile communications/ICT |
|---|---|
| **1. End poverty in all its forms everywhere** (No poverty) | Mobile communications can provide a communication infrastructure to stimulate local economy growth in poor communities; lowers barriers to access economic resources by providing access to mobile money and micro-financing; and generates employment opportunities for people living in extreme poverty [GSMA 2018]. <br> Digital financial services allow many to participate in the digital economy for the first time; access to financial services has proven to be a pivotal step in helping people lead out of poverty; timely and accurate information services will help ensure equal rights to economic resources and market insights that can benefit all [ITU]. |
| **2. End hunger, achieve food security and improved nutrition and promote sustainable agriculture** (Zero hunger) | Mobile communications allow the use of digital agricultural payments, connect remote communities to digital agricultural marketplace, and enhance the efficiency and safety of food production through digitalizing agriculture and agricultural value chains with sensors [GSMA 2018; GSMA 2019]. <br> ICTs help farmers improve crop yields and business productivity through better access to market information, weather forecasts, training programs, and other tailored online content [ITU]. |
| **3. Ensure healthy lives and promote well-being for all at all ages** (Good health and well-being) | Mobile communications enable communication with medical practitioners, monitor well-being through mobile, provide access to health programs, provide digital identity service to access healthcare, improve water quality through IoT, and provides big data for epidemics [GSMA 2018]. It allows access to digital medical records, remote patient monitoring and AR/VR for medical training [Huawei 2019]. <br> Patients can contact health care services remotely regardless of their proximity to a healthcare centre; health care workers can learn and prepare for disease outbreaks, identify patient symptoms, follow established treatment protocols, perform remote diagnostics, and access expert support; big data analytics can help produce snapshots, analyze trends, and make projections on disease outbreaks, health service usage, and patient knowledge, attitudes, and practices [ITU]. |
| **4. Ensure inclusive and equitable quality education and promote lifelong learning opportunities for all** (Quality education) | Mobile communications improve the quality of teaching and learning and facilitate reading and enhanced literacy, and allow access to e-learning and digital materials [GSMA 2018; GSMA 2019]. It facilitates remote learning programs, promotes access to downloadable digital content for rural/remote areas facilitates connected classrooms, provides AR/VR for hands-on learning, facilitates access to massive open online courses (MOOC), provides big data analytics and AI for individualized learning and digital platforms to reach more students per teacher [Huawei 2019]. <br> ICTs power a revolution in digital learning; mobile devices allow students to access learning assets anytime, anywhere; teachers use mobile devices for everything from literacy and numerical training to interactive tutoring; mobile learning has the ability to help break down economic barriers, divides between rural and urban, as well as the gender divide [ITU]. |
| **5. Achieve gender equality and empower all women and girls** (Gender equality) | Use of mobile phone can help women in low- and middle-income countries feel safer and more connected, and provide access to information, services and life-enhancing opportunities [GSMA 2019]. Mobile can connect women to sharing economy and allow access to female-specific (e-health) services [GSMA 2018]. ICT offers digital education platforms for girls/women, mobile-enabled micro-lending to access financial resources, mobile apps and social media through connectivity, and training for digital literacy skills [Huawei 2019]. <br> ICT can enabling everyone to have access to the same online resources and opportunities; ICT enables women to gain a stronger voice in their communities, their government and at the global level; ICTs can also provide new opportunities for women's economic empowerment by creating |

| | business and employment opportunities for women as owners and managers of ICT-accessed projects, as well as employees of new business ventures [ITU]. |
|---|---|
| **6. Ensure availability and sustainable management of water and sanitation for all** (Clean water and sanitation) | New water and sanitation service delivery models are possible using mobile communications, and IoT can be used to improve water monitoring and increase water efficiency and for water quality reporting [GSMA 2018]. <br> ICT enables smart water management, facilitating the measurement and monitoring of water supplies as well as necessary interventions, and enabling practitioners at the local level to ensure the equitable and sustainable extension of water, sanitation and hygiene services; ICT can be integrated into monitoring and evaluation frameworks to optimize operations [ITU]. |
| **7. Ensure access to affordable, reliable, sustainable and modern energy for all** (Affordable and clean energy) | Mobile communications enable access to clean energy solutions through mobile-enabled energy models; pay as you go solar energy service models; and increase proportion of renewable energy used to operate infrastructure; use IoT for energy efficiency [GSMA 2018]. Mobile can help via providing data analytics to determine viability of renewable energy sources; online resource for clean energy procurement; smart grid solutions; blockchain facilitating distributed energy systems; and smart metering [Huawei 2019]. <br> ICT and energy efficiency are connected in two ways: 'Greening of ICTs' and 'Greening through ICTs'. In the first case, ICT is developed to be more environmentally sound and less carbon-intensive. In the second case, ICT-enabled solutions (e.g., smart grids, smart buildings, smart logistics and industrial processes) help to transform the world towards a more sustainable and energy efficient future and can significantly reduce global greenhouse gas emissions [ITU]. |
| **8. Promote sustained, inclusive and sustainable economic growth, full and productive employment and decent work for all** (Decent work and economic growth) | Mobile communications can increase market size by online channels with consumer connectivity and provide access to mobile financial services for companies [GSMA 2018]. <br> ICT skills are a prerequisite for almost all forms of employment. ICT transforms the way that business is being done everywhere and creating new employment opportunities [ITU]. |
| **9. Build resilient infrastructure, promote inclusive and sustainable industrialization and foster innovation** (Industry, innovation and infrastructure) | Mobile networks are a critical infrastructure to provide affordable access to voice and data services and connect remote areas to employment opportunities. IoT solutions can be used for sustainable manufacturing. Mobile infrastructure needs to be developed in a sustainable manner (energy efficiency, community power concept) [GSMA 2018; GSMA 2019]. <br> ICT provides industrial IoT solutions for manufacturing and logistics; smart metering; and smart building automation systems [Huawei 2019]. <br> Industrial grade private networks will be of special relevance[4]. <br> Broadband is essential infrastructure due to its capacity to power industry and innovation; globally harmonized spectrum and standards are essential to facilitate the development of transformative digital infrastructure, such as 5G systems, that will drive scalable solutions to all SDGs [ITU]. |
| **10. Reduce inequality within and among countries** (Reduced inequalities) | Mobile communications enable access to information/social networks to promote social and political inclusion, allows access to marketplaces, and facilitates mobile money and digital identity services [GSMA 2018]. <br> ICT enables access to information and knowledge to disadvantaged segments of society – including those living with disabilities, as well as women and girls [ITU]. |
| **11. Make cities and human settlements inclusive, safe, resilient and sustainable** (Sustainable cities and communities) | Mobile technology can support disaster response, monitor environment through IoT and big data, prevent e-waste, use IoT for municipal planning, and improve transportation efficiency through IoT [GSMA 2018]. <br> ICT provides smart city solutions, e-government initiatives, public transport and improved mobility through online information (smart mobility) and intelligent traffic control systems [Huawei 2019]. <br> ICT offers innovative approaches to managing cities more effectively and holistically – through applications such as smart buildings, smart water management, intelligent transport systems, and new efficiencies in energy consumption and waste management. ICT can be used to make cities more eco-friendly and sustainable [ITU]. <br> Mobile communications can accelerate service creation by enabling continuous innovation, break down silos that limit the impact of new services, engage citizens and stakeholders more directly, and stretch limited resources with enablers such as broadband networks and digital platforms[5] <br> Advanced communications techniques can attract and retain workers in rural areas by improving the teleworking experience, allowing for collaborative innovation systems among firms and |

---

[4] https://pf.content.nokia.com/t004f8-why-private-wireless/white-paper-industrial-grade-private-wireless-for-industry-4-0-applications

[5] https://pages.nokia.com/T003OK-city-as-a-platform.html

| | |
|---|---|
| | research centers, and increasing efficiency of rural business and training of workers. In addition, self-driving cars can improve public transport and increase attractiveness to live and work in rural areas while drones can boost productivity of rural business, improve access to goods, and reduce productions and delivery costs [OECD 2019a, p. 15]. |
| **12. Ensure sustainable consumption and production patterns (Responsible consumption and production)** | Mobile communications can increase energy efficiency of operating infrastructure, increase the proportion of clean energy in value chain, implement best practice waste management, improve awareness by providing access to information through mobile and provide IT solutions to improve sustainability [GSMA 2018]. <br> Increased dematerialization and virtualization as well as innovative ICT applications enable sustainable production and consumption. Cloud computing, smart grids, smart metering, and reduced energy consumption of ICTs have a positive impact on reducing consumption. ICT energy consumption and negative impacts of ICTs, such as e-waste, need to be minimized [ITU]. |
| **13. Take urgent action to combat climate change and its impacts (Climate action)** | Mobile communication can help to avoid greenhouse gas (GHG) as emissions by enabling other sectors to reduce their GHG emissions, facilitate smart traffic management, smart urban lighting, smart parking, smart logistics, building energy management systems, remote working, sharing economy, smart grids, connected health, and precision agriculture [GSMA 2019]. It can improve resilience to the effects of climate change by predicting climate disasters, alerting high-risk citizens, and supporting vulnerable communities [GSMA 2019]. It can also reduce emissions, driving energy efficiency of networks, sourcing renewable energy, decreasing value chain emissions in mobile sector [GSMA 2019]. <br> ICT plays a crucial role in earth monitoring, sharing climate and weather information, forecasting, and early warning systems enabling both the global monitoring of climate change as well as strengthening resilience by helping mitigate the effects of climate change through forecasting and early warning systems [ITU]. |
| **14. Conserve and sustainably use the oceans, seas and marine resources for sustainable development (Life below water)** | IoT solutions and infrastructure can monitor and manage coastal marine ecosystems, connect fishing communities, collect data and share it to improve sustainability of marine environment [GSMA 2018]. <br> ICT facilitates conservation and sustainable use of the oceans through improved monitoring and reporting delivering timely and accurate data on a global basis, while local sensors deliver on the spot updates in real-time. Big data can be used to analyse short- and long-term trends in terms of biodiversity, pollution, weather patterns and ecosystem evolution, and to plan mitigation activities [ITU]. <br> Enabling technologies in the ocean economy can improve data quality, data volumes, connectivity and communication from the depths of the sea. Blockchain and big data analytics applications are being deployed in port facilities and maritime supply chains. Autonomous ships and autonomous underwater vehicles (AUVs) [OECD 2019b, p. 48]. |
| **15. Protect, restore and promote sustainable use of terrestrial ecosystems, sustainably manage forests, combat desertification, and halt and reverse land degradation and halt biodiversity loss (Life on land)** | IoT and wireless support forest monitoring, monitoring of mountain ecosystems and effectively dispose of operational waste [GSMA 2018]. <br> ICT allows conservation and sustainable use of terrestrial ecosystems and the prevention of the loss of biodiversity through improved monitoring and reporting by delivering timely and accurate data on a global basis, while local sensors can deliver on the spot updates in real-time. Big data can be used to analyse short- and long-term trends in terms of biodiversity, pollution, weather patterns and ecosystem evolution, and to plan mitigation activities [ITU]. |
| **16. Promote peaceful and inclusive societies for sustainable development, provide access to justice for all and build effective, accountable and inclusive institutions at all levels (Peace, justice and strong institutions)** | Mobile technology for authorities (e.g., police) helps to prevent violence and adheres strict data privacy and security policies that align to national and international law [GSMA 2018]. <br> ICT helps in crisis management, humanitarian aid and peacebuilding. Growing use of open data by governments increases transparency, empowers citizens, and helps to drive economic growth by record-keeping and tracking government data and local demographics. In natural or man-made disasters ICT can help in obtaining, communicating and transmitting accurate and timely information. big data analysis and data mining can make use of vast amount of data accessible online [ITU]. |
| **17. Strengthen the means of implementation and revitalize the Global Partnership for Sustainable Development (Partnerships for the goals)** | Mobile sector supports collaboration between public and private sector [GSMA 2019]. <br> ICT is crucial in achieving all of the SDGs, since ICT acts as catalysts that accelerate all three pillars of sustainable development – economic growth, social inclusion and environmental sustainability – as well as providing an innovative and effective means of implementation in today's inter-connected world [ITU]. |

Table 4-1 clearly brings out that affordable access to internet by everybody is critical to achieving the SDG goals as access to information is the panacea for everything and is significantly cheaper and faster than for example building physical infrastructures of schools, hospitals, etc.; all infrastructures providing services that can be partly offered remotely with the help of internet connection and appropriate applications.

### 4.1.2 Existing indicators

In the global indicator framework for the UN SDGs [UN 2017], there are seven indicators that are defined to be related to ICT and are systematically monitored. They are highlighted in bold in Table 4-2. These indicators present a very high-level view and are not adequate for investigating the impact of mobile communications on the UN SDGs. In order to provide a more holistic view on what can and even should be measured we need other metrics as well. The metrics can serve various stakeholders in the mobile communication industry to guide their activities into a more sustainable direction, in a more sustainable manner. Traditionally, cellular mobile communication networks have been characterized by sets of minimum technical requirements for the different generations of International Mobile Telecommunications (IMT) systems defined by the ITU. The latest systems, IMT-2020, need to fulfill requirements on the following indicators: peak data rate, peak spectral efficiency, user experienced data rate, 5th percentile user spectral efficiency, average spectral efficiency, area traffic capacity, latency in terms of user plane latency and control plane latency, connection density, energy efficiency, reliability, mobility, mobility interruption time and bandwidth as illustrated in Figure 6-1 [ITU-R 2017]. With the advent of 6G, a new set of indicators needs to be defined.

The impact of mobile communications on the UN SDGs has been measured with indicators which are summarized in Table 4-2 where the bolded indicators are from the UN Global Indicator Framework for the SDGs.

**Table 4-2.** Existing indicators related to UN SDGs and mobile communications/ICT.

| SDGs | Existing indicators related UN SDGs and mobile communications/ICT |
|---|---|
| **1. No poverty** | 2G coverage; 3G coverage; 4G coverage; mobile penetration of the poorest 40%; mobile money penetration; number of transactions per account; average transaction volume; mobile money registered accounts [GSMA 2018]. <br> Proportion of individual using the Internet; proportion of households with Internet access; proportion of individuals owning a mobile phone; proportion of individuals using the Internet for Internet banking [WSIS 2019]. |
| **2. Zero hunger** | Mobile penetration; receiving of payments for agricultural products via mobile; number of people using mobile access that benefit their farm or fishery; use of mobile to access health services [GSMA 2018]. <br> Proportion of individual using the Internet; proportion of individuals owning a mobile phone; population covered by a mobile broadband network [WSIS 2019]. |
| **3. Good health and well-being** | Proportion of individuals using a mobile phone; countries having adopted a national e-health record [WSIS 2019]. |
| **4. Quality education** | **Proportion of schools with access to the Internet for pedagogical purposes; proportion of schools with access to computers for pedagogical purposes; proportion of youth/adults with ICT skills, by type of skills [UN 2017].** <br> Individuals with ICT skills by type of skill by age; percentage of youths/adults who have achieved at least a minimum level of proficiency in digital literacy skills; proportion of individuals using the Internet; enrolment in basic computer skills and/or computing courses in secondary education; proportion of graduates in ICT-related fields as post-secondary levels; proportion of educational institutions with computers for pedagogical purposes; proportion of educational institutes with Internet for pedagogical purposes; learner-to-computer ratio [WSIS 2019]. |
| **5. Gender equality** | **Proportion of individuals who own a mobile telephone, by sex [UN 2017].** <br> Proportion of individuals using the Internet [WSIS 2019]. |
| **6. Clean water and sanitation** | Total access gap (water, sanitation) [GSMA 2018]. |

| 7. Affordable and clean energy | Number of people benefitted from access to clean and reliable energy in their homes; IoT utilities connections; number of people in developing countries using mobile to pay utility bills; energy use by business area of MNO [GSMA 2018]. |
|---|---|
| 8. Decent work and economic growth | Population covered by a mobile broadband network; individuals with ICT skills by type of skill by age; internet traffic in exabytes; Business' use of broadband subscriptions; International trade in digitally-delivered services as a percentage of total services trade %; businesses using the Internet for Internet banking for accessing other financial services; proportion of individuals using the Internet for Internet banking; proportion of e-waste treated environmentally sound [WSIS 2019]. |
| 9. Industry, innovation and infrastructure | **Percentage of the population covered by a mobile network, broken down by technology [UN 2017].** <br> Mobile subscribers; mobile internet adoption; 2G coverage; 3G coverage; 4G coverage; cost of 500 MB data plan; average download speed; average upload speed; average latency [GSMA 2018]. Proportion of households with Internet access; population covered by a mobile broadband network; educational institutes (schools) with Internet; ICT prices as a % of gross national income; international Internet bandwidth (bps) per Internet user [WSIS 2019]. |
| 10. Reduced inequalities | Mobile penetration for the poorest 40% of population in developing countries; mobile money adoption among women in low-income countries; percentage of refugees living in areas covered by mobile networks, percentage of refugee households having a mobile phone, number of mobile money transactions, affordability of sending mobile money [GSMA 2018]. |
| 11. Sustainable cities and communities | IoT smart city connections; IoT smart vehicle connections; number of operators disclosing waste information [GSMA 2018]. |
| 12. Responsible consumption and production | Proportion of e-waste treated environmentally sound; Proportion of individuals using the Internet [WSIS 2019]. |
| 13. Climate action | |
| 14. Life below water | |
| 15. Life on land | |
| 16. Peace, justice and strong institutions | United Nations' E-participation index; proportion of individuals using the Internet; proportion of individuals owning a mobile phone [WSIS 2019]. |
| 17. Partnerships for the goals | **Fixed Internet broadband subscriptions, broken down by speed; proportion of individuals using the Internet [UN 2017].** <br> Proportion of businesses using the Internet; proportion of businesses receiving orders via the Internet; proportion of businesses placing orders over the Internet [WSIS 2019]. |

### 4.1.3 Identifying opportunities of 6G

Building on the existing mapping between the UN SDGs and mobile communications or ICT, we note that future technologies, such as 6G, are not thoroughly captured in the linkages or indicators. The UN SDGs can be seen as an opportunity for 6G to make a positive impact and be the accelerator for the advancement of the UN SDGs. Reducing inequalities and securing a better quality of life for those in the most vulnerable position in a sustainable way is at the heart of the goals. The pursuing of the UN SDGs is a social and societal effort and can direct public resources to, among other things, extending the network infrastructure required for 6G beyond urban centers and developed countries. A prerequisite for this is the creation of 6G use cases that support the UN SDGs and can guide the development of the standard. Implementing the 6G technology should be affordable and not require disproportionate infrastructure investment. Another opportunity is to exploit the existing and evolving infrastructure is already being build including fiber optic backhaul infrastructure. An overview of the opportunities presented by 6G in assisting the achievement of UN SDGs identified in the White Paper Expert Group is summarized in Table 4-3 using the PESTLE analysis framework.

**Table 4-3.** Identified key opportunities of 6G.

| Dimension/ Perspective | Opportunities |
|---|---|
| **P - Political** | <ul><li>Communication networks are seen as a necessity for basic standards of living.</li><li>Increasing consensus on funding networks in areas with low business opportunity.</li><li>Users can manage the most personal data, even if web scale companies still own masses of our data.</li></ul> |

| Dimension/ Perspective | Opportunities |
|---|---|
| **E - Economical** | - Better communication infrastructure can stimulate local economy growth in poor communities by lowering barriers to access economic resources by supporting access to financial services and generating employment opportunities for people living in extreme poverty.
- 6G may facilitate small businesses through increased market transparency: helping to build trust in online services and supporting cross-border e-commerce through better (online) business security and data protection
- 6G may help in generating and finding employment opportunities, supporting access to financial services, facilitating access to utility services, providing digital identity services and emergency and disaster services
- Opportunities for entrepreneurship or coops with connectivity and online mentoring.
- Access to reliable online weather forecasts to reduce unnecessary watering and to define optimal cultivation time. Affordable analysis of soil quality benefitting of AI.
- Data collection with online tools and support for identifying breeds e.g. with shape recognition. Also, peer-support and networking possibilities to reduce inbreeding in cattle. |
| **S - Societal** | - Mobile communications can help with illiteracy and if not that, still help in reducing inequalities due to illiteracy.
- Having easy online access to land ownership or property data with possibilities to also sign and verify documents' authenticity with e.g. blockchain can revolutionize ownership for the most vulnerable.
- Remote consultations for pregnant women including measuring unborn child's heartbeat. For risk pregnancies, e.g. mothers expecting twins, ultrasound to be replaced with novel low-cost technologies with aid of safe in-body measurement devices or similar.
- Leverage 6G for smart water grid management/6G for remote water resource management and supervision.
- Emission trade based on globally distributed sensors measuring real time emissions. |
| **T - Technological** | - The access to free (or affordable) and virtually unlimited spectrum resources is among the key factors that influences the capability of mobile communications technologies to contribute to the UN SDGs. In 6G, the utilization of spectrum sub-THz and THz bands above 100 GHz may significantly relax spectrum scarcity constraints commonly encountered in preceding generations. Moreover, these THz not only support scalability of communications, but may also be conveniently exploited for high-precision localization, sensing and imaging applications even in underserved ('under-sensed') areas.
- The adoption of open source principles and/or a unified global standard and early interaction with all stakeholders in development of mobile technologies, platforms and services may further unlock opportunities for co-innovation and co-creation within the ecosystem, with possibility of more meaningful involvement of the UN SDGs beneficiary stakeholders. The confluence of mobile technology development and open source models is becoming increasingly prominent in 5G and will likely be the norm in the 6G era.
- The technical KPI enhancements (e.g. faster data rates, lower latencies etc.) envisioned in 6G (relative to 5G KPIs) may potentially bring deeper immersion and interaction for users (e.g. using holographic interactions, tactile Internet etc.), even in hitherto underserved areas. Such capabilities may add value to remote service delivery (e.g. health, education etc.) in ways that far exceeds tele-X solutions from legacy mobile technologies.
- The envisioned migration of most computing and intelligence to the edge (closer to end users) in 6G may further contribute to bringing technology-enhanced SDG interventions closer to local points of need. To that end, 6G contributes in way that enhances service immediacy, local autonomy and adoption to local environmental contexts (both physical and virtual). |
| **L - Legal and regulatory** | - Sustainability for driving spectrum use to ensure efficient use of spectrum and supporting different stakeholders to allow the deployment of 6G networks |
| **E - Environmental** | - 6G has potential to enable solutions in various disciplines, e.g. virtual learning, smart travelling etc., that eventually decrease the carbon footprint.
- If energy needs of the 6G technology will be provided by low-cost renewable energy technologies, it will promote the use of clean and affordable energy technologies.
- The climate change goal to limit the goal warming by 2 degrees from pre-industrial era can be facilitated by using a combination of 6G technology together with clean energy technology.
- Efficient and intelligent 6G network could improve the communication and energy systems resilience and could play a key role in achieving zero (or even negative) carbon emission solutions.
- Innovative solutions to promote circular economy are possible using the 6G technology. |

## 4.2 Developing a novel linkage between 6G and the UN SDGs via indicators

This White Paper set out to establish the mapping of 6G on the UN SDGs, a task that has proven to be rather elusive, because there is no other prior work on the topic. To address this gap, this White Paper Expert Group carefully examined the possible linkages between 6G and the UN SDGs, leading to the following ex-ante three tier analytical framework that was preliminarily introduced in Chapter 1. Specifically, the proposed framework draws linkages to the UN SDGS by viewing 6G as: **1) a provider of services to help steer communities and countries towards reaching the SDGs', 2) an enabler of measuring tool for data collection to help reporting of indicators with hyperlocal granularity, and 3) a reinforcer of a new technological ecosystem to be developed in line with the UN SDGs**. Next we briefly expand on the proposed three pillars.

### 4.2.1 6G as a provider of services

The 2030 Agenda sets forth the blueprint for tackling focal problems facing society today across the globe. It becomes evident that irrespective of geographical location or geopolitical standing, many communities become progressively worse while being trapped into downward feedback spirals that prevent people, authorities, and business reach their goals in a sustainable manner. For the most part, the lack of visible and viable opportunities, that could otherwise facilitate people and institutions steer their actions towards community betterment, cultivate meta-narratives of hate, deviance, polarisation, environmental degradation, radicalisation, conflict, terrorism, and war. For example, one can observe that across the Human Development Index (HDI), communities lacking access to good education, cannot contribute positively to the development of job and wealth creation that leads to social change and advancement. In this context, 6G technology will deliver services and solutions that will empower individuals and communities to adopt self-correcting processes and steer the actions towards long term sustainability.

Specifically, next-generation 6G wireless services built on pervasive Artificial Intelligence and fractal network protocol stacks will have a profound effect on how individuals and communities perceive *space* and *time*. For instance, multisensory applications (VR, AR, XR, MR) will bring about new forms of human mobilities that will enable individuals and large groups to break away from existing patterns of interaction and temporo-spatial understandings of the world. New smart surfaces, autonomous systems, and wireless brain-computer interactions will also reshape our perception of time and space in drastically new ways.

### 4.2.2 6G as an enabler of hyperlocal measuring tool

Following the initial publication of the UN SDG indicators [UN 2017], many international governmental and non-governmental organisations voiced concern about existing data gaps that hinder the capacity of authorities to report on their performance. Today, of the 169 targets in the Agenda, OECD countries report data for only 105[6]. The White Paper Expert Group examined further the scope of this data-gap and conducted a systematic analysis of all available UN SDG indicators metadata, looking at how indicators are measured in the field[7]. The conclusions of this exhaustive exercise are twofold: (1) most indicators are currently being measured with asynchronous and anachronistic tools (questionnaires or household surveys) and are not up to par with current advances in data management and data retrieval; (2) there is a discord between the breadth of some indicators and the inefficacy of instruments to collect disaggregated data. Against this backdrop, this White Paper anticipates the *instrumentation* of 6G as a scientific tool for helping authorities and other stakeholders to collect data at a much more local, hyperlocal and granular level than currently conducted. The use of 6G technology as a measurement tool will not only enable the reporting of missing KPIs, and thus, reducing the data-gap, but will also enable the monitoring of new indicators that will become relevant in the near future. The expert group has identified a four-layer taxonomy comprised of cross-cutting domains that can be used for eliciting more refined indicators for the UN SDGs': Critical infrastructure, ICT/mobile network vendors, marketplaces, and application services.

---

[6] http://www.oecd.org/development/launch-of-measuring-distance-to-the-sdg-target-may-2019.htm
[7] https://unstats.un.org/sdgs/metadata/

For instance, mobile coverage is a basic prerequisite for the utilisation of the IoT ecosystem that help communities meet their goals. Sensors that capture and transmit data quickly over WiFi, 3G, 4G and 5G are becoming more accurate and efficient. The development of mobile networks helps data from the sensors to be collected and transmitted more efficiently faster and across larger distances. An exponentially increasing amount of data is made available from all types of geographical, economic, and social contexts, while the technologies for system and sub-system monitoring is growing in fully maturity. Thus, the instrumentation of 6G technology for monitoring of KPI's is expected to have major improvement in increasing the sampling rate, improving the veracity and variety of the measured data, reduce costs. Communication infrastructure (mobile networks) can be used as a sensor and as innovative tool to monitor sustainability and efficiency in many different forms and contexts. For example, in the developed world, the penetration of home IoT devices can be treated as a sign of prosperity and well-being, enabling governments to identify those neighbourhood blocks that require more support than others. Respectively, indicators like volume of streaming content, data consumption, and data production can all lead to better hyperlocal KPIs which can be monitored, reported and addressed by stakeholders in near real-time.

### 4.2.3 6G as a reinforcer of ecosystem aligned with the UN SDGs

It is anticipated that 6G technology will have a disruptive impact that will bring about positive changes in the way individuals, communities and governments steer to solutions across all aspects of life. At the individual level, the envisioned services that will be derived by the 6G wireless systems will call for a new bottom-up mandate of social relations and conduct. For example, hyperlocal monitoring of energy consumption and $CO_2$ footprint can work as a reinforcing feedback loop leading to individual self-adjustment. For example, new services will provide individuals and communities with access to KPIs on their personal energy footprint. The impact of 6G will also be relatively greater in the domain of governance, public administration, and through institutions (e.g. 6G Hospitals, 6G Schools, Industry 4.0, e-government). The interconnection between people, communities, and institutions along with the increasing use of heterogenous data streams in the area of public policy decision-making, is set to reconfigure how people, governments and the industry will be governed in the future. For example, it has often been recognised that the role of 6G as an enabling technology needs to be developed in accordance with the relevant UN SDGs including, but not limited to the new governance paradigm on energy efficiency, recycled materials, non-toxic leakage, inclusiveness, partnerships, privacy/security, and democracy. Thus, high-end novel technology solutions enabled by 6G such as telepresence and mixed reality, accurate positioning, wearable displays, mobile robots and drones, specialized processors, and next-generation wireless networks will raise new questions with regards to law, privacy, values, corporate accountability and government transparency. The first 6G White Paper [6G Flagship 2019] outlined that the current smart phones are likely to be replaced by pervasive XR experiences with lightweight glasses delivering unprecedented resolution, frame rates, and dynamic range. But at the same time, we must ensure that 6G is developed keeping mind the UN SDGs for example to work towards maximum coverage/connectivity also in the rural areas.

### 4.2.4 Linkage between the UN SDGs and 6G

The starting point in finding novel linkages between 6G and the UN SDGs was to see which targets, and more specifically, which indicators 6G can impact. The expert group looked into where 6G can impact the indicators by identifying what 6G can do to influence the indicators [UN 2017] to meet the given UN SDG targets. The expert group used the developed three-tier analytical framework of 6G when examining the linkages between individual targets and their indicators. First a set of goals was selected based on the input of the expert group, then the targets for these goals were examined together with the matching UN indicators, and finally a framework of how 6G can either 1) provide services to help steer communities and countries towards reaching the SDGs, 2) enable measuring for data collection to help reporting of indicators, and 3) reinforce a new technological ecosystem to be developed in line with the UN SDGs. When doing this, the cross-cutting issues (entry points for transformation) as suggested by the UN Global Sustainable Development Report [UN 2019b] were considered: Human well-being and capabilities, Sustainable and just economies, Food systems and nutrition patterns, Energy decarbonization with universal access, Urban and peri-urban development and Global environmental commons which are summarized in Fig. 1-1. Table 4-4 presents examples of the developed linkages per selected goals so that they present each of the entry points. These examples serve as the starting point for a more detailed mapping of 6G with the UN SDGs that we encourage the wider community to work on.

**Table 4-4.** Mapping of 6G with UN SDG framework targets via indicators.

| UN Targets | UN Indicators | 6G can |
|---|---|---|
| **1.2** By 2030, reduce at least by half the proportion of men, women and children of all ages living in poverty in all its dimensions according to national definitions | **1.1.1** Proportion of population below the international poverty line, by sex, age, employment status and geographical location (urban/rural) <br> **1.2.1** Proportion of population living below the national poverty line, by sex and age | ⇒ Expand the connections between people with processes, data, and businesses. <br> ⇒ Help incorporate one-man shops into the market <br> ⇒ Facilitate the use of micro-payments and digital money. <br> ⇒ Provide advanced life-long learning opportunities to previously marginalized areas. <br> ⇒ Help train local educators with remote virtual platforms. <br> ⇒ Maximize the use of ICT by delivering virtual computational power to remote areas. |
| **1.3** Implement nationally appropriate social protection systems and measures for all, including floors, and by 2030 achieve substantial coverage of the poor and the vulnerable | **1.3.1** Proportion of population covered by social protection floors/systems, by sex, distinguishing children, unemployed persons, older persons, persons with disabilities, pregnant women, newborns, work-injury victims and the poor and the vulnerable | ⇒ Achieve better prevention measure by enabling self-risk and self-awareness through remote learning. <br> ⇒ Extend the use of remote eHealth services and diagnostics. <br> ⇒ Create new labor markets for remote rural areas enabling by smart farming and autonomous control systems. <br> ⇒ Bolster the connectivity of production and consumption markets with the use of autonomous vehicles. <br> ⇒ Achieve real time communication infrastructure between institutions (hospitals, schools, governments) to enable better analytics and predictions. |
| **3.3** By 2030, end the epidemics of AIDS, tuberculosis, malaria and neglected tropical diseases and combat hepatitis, water-borne diseases and other communicable diseases | **3.3.1** Number of new HIV infections per 1,000 uninfected population, by sex, age and key populations <br> **3.3.2** Tuberculosis incidence per 100,000 population <br> **3.3.3** Malaria incidence per 1,000 population <br> **3.3.4** Hepatitis B incidence per 100,000 population <br> **3.3.5** Number of people requiring interventions against neglected tropical diseases | ⇒ Provide digital health solutions that will increase self-awareness of risks. <br> ⇒ Deliver eHealth monitoring services (bloodsugar, liver, heart) to mothers and children during and after pregnancy. <br> ⇒ Enable e-registers for monitoring the availability of vaccines. <br> ⇒ Help implement a global early warning alerting mechanism. <br> ⇒ Enable real-time communication between hospitals for sharing of information about positive cases and availability of medical resources. <br> ⇒ Accelerate scientific breakthroughs with increasing sharing of information and computational resources. <br> ⇒ Support remote learning, diagnostics, and training to medical staff and nurses. <br> ⇒ Provide hyperlocal notifications and contact tracing capabilities in regions where electricity and communication networks are unreliable. <br> ⇒ Deploy unmanned vehicle for food and medical deliveries in lockdown areas. |
| **4.2** By 2030, ensure that all girls and boys have access to quality early childhood development, care and pre-primary education so | **4.2.1** Proportion of children under 5 years of age who are developmentally on track in health, learning and psycho-social well-being, by sex | ⇒ Increase access to remote learning and developmental activities to children under 5 years. <br> ⇒ Enable improved socialization through virtual interactions <br> ⇒ Improve remote access to Pediatrics in location with poor connectivity. <br> ⇒ Facilitate remote and virtual training of local pediatricians |
| that they are ready for primary education | 4.2.2 Participation rate in organized learning (one year be-fore the official primary entry age), by sex | ⇒ Help improve and develop the knowledge and skills of local medical community.<br>⇒ Deliver prosthetic technologies to support handicapped children.<br>⇒ Permit family and experts to monitor the cognitive development of children with Brain-Computer Interfaces.<br>⇒ Help coordinate virtual meeting for preschoolers. |
| 7.3 By 2030, increase substantially the share of renewable energy in the global energy mix. | 7.2.1 Renewable energy share in the total final energy consumption<br>7.3.1 Energy intensity measured in terms of primary energy and GDP | ⇒ Help modernize power grids to improve efficiency and sustainability at lower cost (smart grids).<br>⇒ Provide support for autonomous system control with faster distributed measurement technologies.<br>⇒ Enable scheduling and connectivity with domestic appliances. |
| 9.c Significantly increase access to information and communications technology and strive to provide universal and affordable access to the Internet in least developed countries by 2020 | 9.c.1 Proportion of population covered by a mobile network, by technology | ⇒ Help telecommunication operators achieve greater mobile network coverage enabling access also to the most remote areas.<br>⇒ Increase the percentage of inhabitants living within range of mobile-broadband network.<br>⇒ Improve market competition by enabling smaller operators to participate in vertical markets. |
| 11.5 By 2030, significantly reduce the number of deaths and the number of people affected and substantially decrease the direct economic losses relative to global gross domestic product caused by disasters, including water-related disasters, with a focus on protecting the poor and people in vulnerable situations | 11.5.1: Number of deaths, missing persons and directly affected persons attributed to disasters per 100,000 population | ⇒ Facilitate immersive virtual training to front line practitioners.<br>⇒ Implement rapid deployment force of autonomous and unmanned rescue vehicles in disaster areas to maximize the discovery of missing persons.<br>⇒ Help create radio technology to locate trapped victims.<br>⇒ Develop hyperlocal warning and emergency alert systems that prevents deaths.<br>⇒ Improve personal awareness of risks.<br>⇒ Improve connectivity in disaster areas where energy and communication sources are damaged.<br>⇒ Enlarge existing disaster risk reduction strategies by providing remote training to local governments.<br>⇒ Improve national, regional and multinational disaster coordination system. |

### 4.3 Possible 6G Indicators

The White Paper Expert Group produced an extensive list of indicators that either measure the development or the performance of 6G aligned with the UN SDGs, or the value it can provide to the advancement of the sustainable development goals. In order to classify the proposed 6G indicators, we used our three-tier analytical framework from Section 4.2 together with the entry points discussed in Chapter 1 and inserted the suggested indicators in the relevant categories.

It has already become visible that wealth generated by economic growth will become concentrated in the hands of a shrinking minority and that rates of growth will be low in Western countries. Boundaries to growth will be set by for example the ageing population especially in the western countries and our planet's ecological carrying capacity all around the world. This means that statistical indicators of well-being and genuine development will take on a more important role[8]. Based on our work it seems that the proposed indicators for 6G already reflect this vision: most indicators suggested measure the services 6G can provide to enhance human well-being.

---

[8] https://media.sitra.fi/2020/03/04130112/2021544megatrendikortit2020enverkko.pdf

Table 4-5 presents the new 6G related indicators developed within the Expert Group per entry points. The entry points were used in the classification of indicators because they are cross-cutting for all 17 goals and offer a more holistic view. The entry points are described in more detail in [UN 2019b, pp. 163-71] and summarized in the table. Looking at the indicators through the entry points also ensures the creation of a framework where no individual goal is being overly advanced at the cost of another. Having indicators for different entry points allows us to set target values that tackle more than just one goal at a time. We present indicators that can be measured in numbers but there are many fundamental issues such as the availability of a certain 6G enabled service or the coverage that can be measured by simply answering yes/no. The latter should provide the baseline for numeric indicators.

It should be noted that Table 4-5 presents a very preliminary view on possible 6G related indicators that act as the opening of the debate on what kind of indicators should be defined for 6G while taking into account the growing importance of meeting the UN SDGs. They are not necessarily specific to future 6G technologies. In addition, no target values have yet been defined for the indicators. We invite the wider community to develop 6G specific indicators aligned with the UN SDG developments towards 2030.

**Table 4-5. 6G indicators per entry points to UN SDGs.**

| Entry points from [UN 2019b] | 1) 6G as provider of services | 2) 6G as enabler of hyperlocal measuring tool | 3) 6G as reinforcer of ecosystem aligned with SDGs |
|---|---|---|---|
| **Human well-being and capabilities**<br>*A.1. All stakeholders should contribute to eliminate deprivations and build resilience across multiple dimensions through universal provision of, and access to quality basic services (health, education, water, sanitation, energy, disaster risk management, information and communications technology, adequate housing and social protection), that are universally accessible with targeted attention where poverty and vulnerability are concentrated and with special attention to individuals who are most likely to be left behind – women and girls, persons with disabilities, indigenous peoples and others.*<br>*A.2. Governments should ensure equal access to opportunities, end legal and social discrimination and invest in building human capabilities so that all people are empowered and equipped to shape their lives and bring about collective change.* | The availability of appropriate technologies and services related to quality basic services: health, education, water, sanitation, energy, disaster risk management, ICT, adequate housing and social protection (Yes/No) and the number of provided 6G enabled services for example: Number of 6G enabled eHealth services.<br><br>Amount of digital mobile telecommunication education utilizing disruptive technologies.<br><br>Proportion of e-learning material to be used with a mobile device with mobile connection out of necessary learning material; proportion of accessible material (for people with disabilities, in different languages.<br><br>The number of standardized procedures to report via mobile app violence/ harmful practices against girls/ women, human trafficking etc.<br><br>Number of people working remotely.<br><br>Number of 6G holographic telepresence access points per population. | Mobile coverage.<br><br>Mobile area coverage.<br><br>6G transmitting sensor data on: living conditions (temperature, humidity, inhabitants per m2) etc.; natural disasters, diseases (epidemics); mobility of humans & animals; quality of water, air etc.<br><br>6G sensing data using radio waves for sensing purposes. | Price of connectivity.<br>Affordable technologies/ devices.<br>Price of privacy. |
| | The price of connection compared to a commodity (e.g. connection price vs. 1 kg rice). | | |
| **Sustainable and just economies**<br>*A3. Governments, international organizations and the private sector should work to encourage investment that is more strongly aligned to longer-term sustainability pathways and to facilitate disinvestment away from those that are less sustainable.*<br>*A4. All stakeholders should work together to achieve a global decoupling of GDP growth from the overuse of environmental resources, with different starting points that require different approaches across rich, middle-income and poor countries.*<br>*A5. Governments, supported by civil society and the private sector, should promote an upward convergence in living standards and opportunities, accompanied by reduced inequalities in wealth and income, within and across countries.* | Mobile subscription considered a basic need in any given society (yes/no).<br><br>Unrestricted access to internet and mobile services (yes/no).<br><br>Number of mobile birth registrations and personal electronic identification.<br><br>Availability of public mobile services such as tax/license services, census, banking etc.<br><br>Availability of mobile financial transactions (micro payments) (yes/no).<br><br>Amount of mobile services supporting circular economies. | Data subscription models available and their relative proportion.<br><br>Use of data (I.e., what subscription models poor people use, do they even have access to data, who uses the data, only the rich ones?).<br><br>Percentile of mobile devices suited for browsing and rich data consumption (chiefly: screen size and resolution) e.g. browsing/web sites, video consumption, gaming, video calls for work and social purposes.<br><br>Share of value in the mobile ecosystem: distribution of monetary value gained and spent per actor in the mobile ecosystem (infrastructure providers such as electricity, operators; device manufacturers; retailers; remanufacturers; second-hand sellers; end-of life material recyclers; content providers; users; etc.) | Distribution of mobile terminals by price (how many expensive terminals are bought in a country).<br><br>% global coverage covered by 6G.<br><br>Business models of operators (just or not)?<br><br>Circular economies in 6G devices.<br><br>E-waste management of new devices. |
| **Food systems and nutrition patterns**<br>*A6. All stakeholders should work to make substantial changes to existing infrastructure, policies, regulations, norms, and preferences so as to transition towards food and nutrition systems that foster universal good health and eliminate malnutrition while minimizing environmental impact.*<br>*A7. Countries must take responsibility for the entire value chain related to their food consumption so as to improve quality, build resilience and reduce environmental impact, with developed countries supporting sustainable agricultural growth in developing countries.* | Percentage of mobile devices having apps to provide information on nutrition.<br><br>Availability of public mobile services such as weather forecast for farmers.<br><br>Number of food waste in kg's in grocery stores/ restaurants using appropriate mobile apps to circulate food going to be wasted.<br><br>Use of smart digital labels.<br><br>Number of IoT devices for agriculture purposes.<br><br>Percentage of farms that are covered by IoT services. | Number of registered users using apps on weather forecasts, to circulate food.<br><br>Data gathered from the use of smart digital labels.<br><br>Data gathered from sensors used in agriculture (for instance the use of fertilizers). | Number of non-toxic, non-leaking devices.<br><br>New technologies e.g. sensors working in harsh conditions.<br><br>New passing sensing functions (device-free) integrated in the mobile communication systems. |
| **Energy decarbonization with universal access**<br>*A8. All stakeholders must ensure universal access to affordable, reliable and modern energy services through the accelerated implementation of costefficient provision of clean electricity, coupled with making clean-* | Number of mobile apps that help to reduce energy consumption. | Number of sensors monitoring energy consumption. | % reduction of energy consumption kWh. |
| *cooking solutions a top political priority, and moving away from using traditional biomass for cooking. All stakeholders should promote clean, reliable and modern energy sources, including by harnessing the potential of decentralized renewable energy solutions.*<br>*A9. International and national entities and stakeholders must collaborate to reshape the global energy system so that it participates fully towards the implementation of Goal 7 by transitioning to net-zero CO2 emissions by midcentury, so as to meet the goals of the Paris Agreement including by introducing carbon pricing and phasing out fossil fuel subsidies.* | | | % increase of energy efficiency.<br><br>% reduction of energy consumption due to new functions to avoid energy consumption. |
| **Urban and peri-urban development**<br>*A10. National governments should give cities the autonomy and resources to engage in effective, evidence-based and inclusive participatory policymaking with an engaged and informed citizenry.*<br>*A11. National governments and local city authorities, in close collaboration with the private sector, should promote people-centred and pro-poor policies and investments for a liveable city that provides decent, sustainable jobs, sustainable universal access to vital services such as water, transport, energy and sanitation, with effective management of all waste and pollutants. Individuals and communities should also scale up their engagement in advancing sustainable urban development.* | Number of autonomous vehicles.<br><br>Number of seamless transport systems.<br><br>Unrestricted access to internet and mobile services (yes/no). | Area coverage to include the rural areas where agriculture, forestry and tourism are big industries. | Number of operators operating in the region. |
| **Global environmental commons**<br>*A12. Governments, local communities, the private sector and international actors must urgently achieve the necessary transformations for conserving, restoring and sustainably using natural resources while simultaneously achieving the Sustainable Development Goals.*<br>*A13. Governments must accurately assess environmental externalities – in particular those that affect the global environmental commons – and change patterns of use through pricing, transfers, regulation and other instruments.*<br>*A14. Stakeholders must work with the academic community in all disciplines to mobilize, harness and disseminate existing knowledge to accelerate the implementation of the Sustainable Development Goals.* | Coverage of monitoring systems used for protecting fauna and flora.<br><br>Apps to report observations related to environmental changes.<br><br>Coverage of IoT services for inhabitants. | Number of sensors monitoring fauna and flora.<br><br>Number of people/ users reporting their local observations related to environmental changes with mobile devices/ apps.<br><br>Number of homes using a certain defined level of IoT (electricity, water consumption, waste management). | Percentage of recycled materials in new 6G devices.<br><br>Reductions in energy consumption in mobile phones, base stations, batteries.<br><br>Reductions in energy consumption due to 6G mobile networks. |

# 5. Key challenges

The UN SDGs form the blueprint for a better and more sustainable future for all [UN 2019a], and the mobile communications sector will be in a unique position to contribute to their achievement with the on-going deployment of 5G networks and the recently started research and development of 6G. However, we also identify a number of issues that, unless properly addressed, will diminish 6G's potential contribution. Reflecting on the megatrends identified in [Dufva 2020], we can point out a number of challenges that 6G design should address. In revisiting the Quintuple Helix model [Carayannis et al. 2012] for 6G-driven sustainable development presented in Chapter 3, we can identify inherent challenges within each subsystem (see Fig. 5-1) that may contribute to undermine 6G's utility for enhancing the systemic synergy of knowledge, know-how, and innovation that is required for the sustainable development of 6G. In this chapter, the key challenges associated with each subsystem are identified and discussed in more depth.

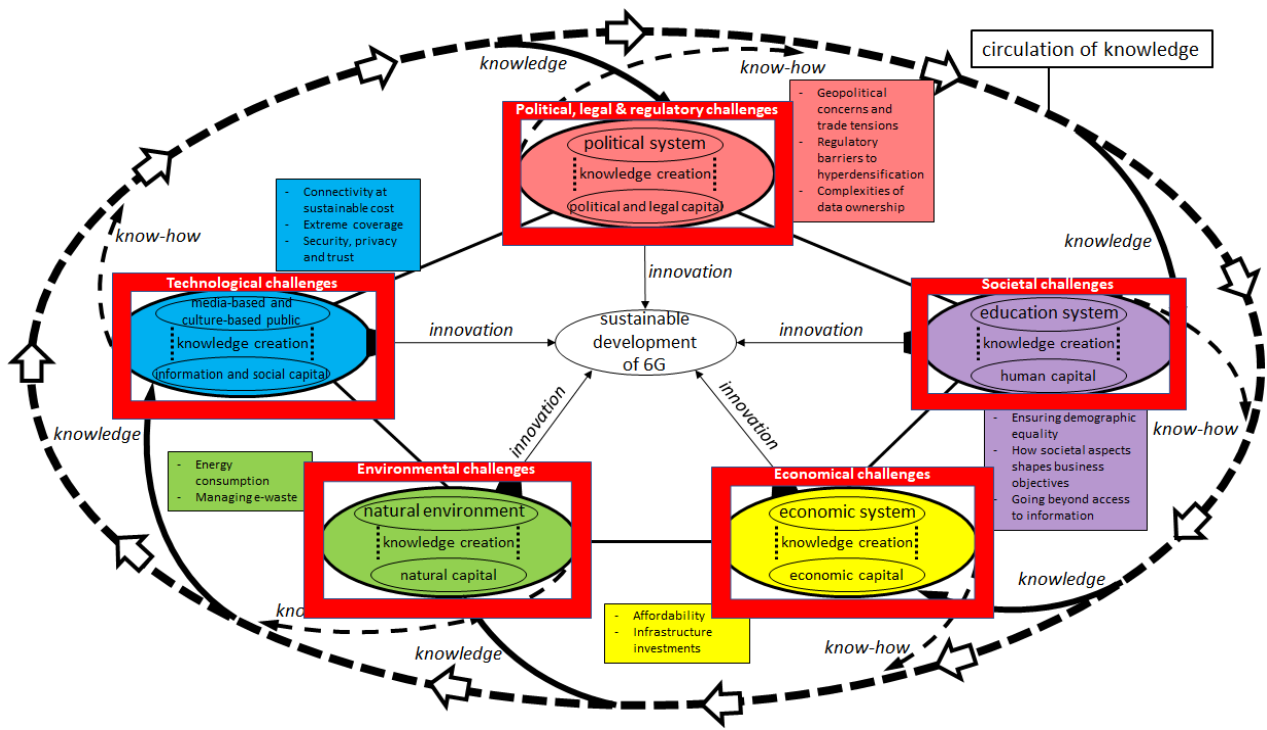

**Figure 5-1.** Identified key challenges within the Quintuple Helix model [Carayannis et al. 2012] for 6G-driven sustainable development.

## 5.1 Political, legal and regulatory challenges

The technology is embedded in political, legal and regulatory systems but it is growingly seen as a geopolitical issue of power raising concerns over the future resource orchestration. Many questions arise including: Who will own the accumulating data? Who will get to decide on technology and who sets the rules and regulation? Spectrum management is at the heart of 6G and any wireless technology development and it will face challenges due to a wide variety of spectrum bands with highly distinct deployment characteristics and spectrum access models with different levels rights and the need for spectrum sharing. The advanced capabilities of 6G will benefit from a fiber backhaul that connects the distributed radio network to the core network. The challenge is that even leading economies have a relatively low fiber penetration, because of underinvestment in pure fiber networks. 6G will likely combine a range of radio access network technologies from macro cells to small cells with very high capacity short-range links, the latter offering denser coverage than 5G and increased capacity and quality of service. Small cell network architectures will be deployed to first and foremost to urban centers and indoors where the natural demand for high bandwidth capacity and low latency is at the highest. Dense small cell deployments have usually been hindered by various factors including regulatory barriers and inconsistent local approval processes[9]. Impact of these inhibitors may be more pronounced for the required hyperdense small cell deployments in 6G.

On the other hand, 6G technology could bring a regulatory disruption allowing advanced specialized networks in a network-of-networks topology. Such sub-networks could be operated by various municipalities, communities or enterprises. Some indications of this are already visible when cities are investing increasingly more on fiber optical infra to enable communication of sensors, which they have installed in various properties and in traffic infra.

## 5.2 Economical challenges

When analyzing the scenarios of redefinition of the economy, a variety of factors supporting the achievement of the UN SDGs are encouraging. Open value configuration and open source principles may bring out high impacts. Similarly, crowdsourcing and crowdfunding may enable unprecedented innovations and industry 5.0 allows to expect revolutionary things. However, there are concerns on how the anticipated novel business models are realized. Given the global trend on the impact of platform, a common concern is how local societies can be empowered.

**Affordability** will be another challenge for 6G as the relative power strengthening of relative power, towards 2030 power configuration is transforming from a multipolar world to a poly-nodal world resulting that power will be determined in the economic, technological and cultural networks and interaction. Previously public network funding has been allocated to areas with low or no coverage, which has recently been changed to favor e.g. smart city development, worksites and ecosystems. There is a real risk that the digital divide will only grow between regions unless the 6G family of standards does not allow for affordable solutions, and address how barriers to digitalization can be overcome. Aspects here include community networks, a freemium model for access and information spots in rural areas to empower the local community.

Subsequent to telecommunications market liberalization in the 1990s, mobile communication infrastructure deployment has been driven from business perspectives, which has ensured the competition and lower prices. On the other hand, the communication infrastructure should be revisited as a natural monopoly like any other essential infra like sanitation or energy having municipal governance. In addition to being fully controlled by commercial operators, the mobile services could be also operated by municipalities or local communities, which could favor more equal access to the services. Naturally this assumes that the instruments for funding the infrastructure investments exist. However, the investment would pay back in societal development by improvement of education, health care, safety or digitalized agriculture and industry. As can be learnt from the underperforming telco natural monopolies in the 60s, however, superior approach may be to frame regulation in the right way while continuing with private sector approach in line with such framework.

## 5.3 Societal challenges

The megatrend addressing digital empowerment and diversification of the population anticipates that learning capabilities will be in key role in the future, which assumes to revisit educational structures and policies. Even if the educational methods will take a huge leap with holopresence, AI and digital twin, high technology is not available equally to all who need it. The trends in mobile only next billion will take monumental efforts and significant change of paradigms.

---

[9] Global5G.org white paper on Small Cells - How Europe can accelerate network densification for the 5G Era, July 2019. http://global5g.org/small-cells

Mere access to information is not enough, rather technology should help people towards acting and collaborating towards achieving sustainability goals in unison. To trigger and prioritize the actions, a change in mindset is needed from business only to business objectives framed by and consistent with societal objectives. Many times, mobile communication has been considered even abundant need, but today after five decades of mobile technology there is plenty of evidence that mobile technology is a pervasive enabler of basic human needs and assuring better standard of life.

Ensuring, **demographic equality** is another significant societal challenge**:** Gender gaps in literacy and political power, and lack of diversity in the composition of innovation teams may harm development of 6G use cases for attainment of the sustainable development goals. Ensuring native security and data privacy is also a key challenge in 6G that calls for global standards and regulations to jointly work on. Advanced technologies should be considered as the network architectures evolve in the future.

## 5.4 Technological challenges

To address issues related to sustainability we may need to work from both dimensions of technological capability and of service attributes such as coverage, latency and reliability. In many cases a quite basic service delivered such as >100 MB/s broadband access to the entire world's population could boost the potential for business, education, health, and further. For these cases the challenge boils down to providing connectivity at sustainable cost, which is a big challenge but not a new one. It can be addressed by several new aspects of 6G: integration of new accesses such as satellites and High-Altitude Platforms (HAP), cost efficient transport and low-consumption nodes. Other challenges can better be addressed by new specific technological components with a more dedicated coverage, for instance in homes, on streets, industries, etc. This could include new spectrum for access, new advanced antenna techniques, precise localization and sensing. Many solutions would be on an intermediate level on both scales, meaning new capabilities with a wide service coverage, such as advanced broadband provided in populated regions of the world. Global sensors would represent an extreme on both scales, requiring both extremely low consumption and extreme coverage.

The 6G era envisions device densities of 1000 connected devices per human or 100 devices per m$^3$ [6G Flagship, p. 14]. Every connected device in a hyperlocal, local community or city context presents a potential entry point to our most sensitive and personal data. Th continuously evolving hyperconnected landscape will expand beyond the contemporary networks in use today. These developments will attract new types of threats and increase complexity of security, privacy and trust challenges, and respective mechanisms to counter the threats.

## 5.5 Environmental challenges

While technology is increasingly embedded in everything, resources should be used more sparingly (ecological reconstruction) and shared more fairly to close up the sustainability gaps. This calls for international consensus at the time when relational power is strengthening and becoming more diverse.

What comes to ecological reconstruction, **energy consumption** in all forms will be a key issue for 6G. While technology offers new solutions for conserving energy, increasing capabilities of technology and increasing use of technology also simultaneously increase the demand for energy. Therefore, the new standard should seek step-change improvement in performance beyond to that achieved by 5G. A ten-fold increase in broadband capacity calls for a comparable increase in energy efficiency per bit to make 6G carbon neutral. Furthermore, the anticipated 6G hyperdense network deployments and device densities would further exacerbate an already critical **e-waste management** challenge, necessitating more radical circular approaches.

# 6. Preliminary action plan for different stakeholders

To truly succeed in achieving the UN SDGs, the mobile communication sector needs to undergo major paradigm shifts to enable affordable access to broadband communication almost everywhere for almost anyone. The paradigm shifts and change in the thought process on our way to 6G needs to happen for all the stakeholders identified in Chapter 2: research and educational organizations; governmental, regulatory and standardization organizations; users; industry including e.g. mobile network operators, network equipment manufacturers, and application and service providers; verticals, and other business drivers. The motto that should drive all the stakeholders is: **"How to provide more for less for more people in a sustainable manner?"** In short, how to make the next generation of wireless affordable without diluting the performance requirements while including the new breed of social impact indicators which are not always directly quantifiable in terms of money. [10]

Pure business driven operations where infrastructure, access and services are primarily dependent on the mobile network operators' business decisions have resulted in situations where the challenge areas with less revenue potential still remain underserved. Therefore, the achievement of the UN SDGs calls for new societal models including, e.g., community driven networks whose emergence is dependent on the underlying regulatory environment. These challenge areas exist even in developed countries where the gap between rural and urban areas is not yet bridged. For example, one critical point is the availability of spectrum which is in the hands of regulators. Concrete actions are needed in making spectrum available for networks in a flexible manner and not only to the existing MNOs. Different actions are needed for rural and urban challenges which are discussed next.

**Research and educational organizations** – The global launch of the 6G research originates from the research community[11] and the need for the research community to continue driving the 6G research is evident. Unbiased research and facilitation of stakeholder interactions for the development of 6G that are not purely commercially driven are important especially in the early stages of 6G. Novel ideas and out-of-the box thinking on various aspects of 6G need to flourish and to allow broad discussions on the most promising and even surprising combinations of technology components needed to make 6G a reality.

**Governments, including Policy Makers and Regulators -** Governments need to lead from the front and in a pro-active manner with long-term visions of the role of ICT and more specifically future 6G in achieving the UN SDGs and to formulate policies and regulations to get there instead of being reactive. For example, formulating policies that will give tail wind to locally deployed networks by different stakeholders, moving away from one regulation one country approach to multiple regimes for the same country optimized for different target user groups and regions, inter-operator roaming solutions that encourage local and private networks, free or low-cost use of the radio spectrum to provide coverage to rural areas, and more revolutionary out-of-the-box thinking to not just auctioning nationwide spectrum license to raise income to the states but to make it available in alternative and shared ways as it is a public and nature's resource, like light, air and water. This would help to bring down the cost of service for everyone. In short, public policies and regulations with long-term objectives must drive the business and deployment of mobile networks, rather than vice versa.

**Standard Developers -** Traditionally, performance requirements and applications have driven the development of every generation of wireless networks and societal impact has been an afterthought. This should change as the industry embarks on developing 6G standards. The goals to achieving the UN SDGs should first and foremost be in developing standards, and as specifically stated in the first 6G white paper [6G Flagship 2019], such as it should be about societal impact and well-being, and by including new indicators. Inclusion of crowd sourcing and user involvement (i.e., communities) should be an integral part of the future open standards development process - not just technology experts.

---

[10] https://www.accessnow.org/cant-reach-u-n-goals-sustainable-development-without-internet/
[11] www.6gflagship.com

**Users -** We expect massive expansion of human possibilities from the way humans will be able to act and interact across physical, digital and biological worlds. Transhumanism trend reflects the rise of technology-driven evolution at an unprecedented speed of change, propelling deeper questions into what it is to be human from biological, behavioral, and human-machine-interfacing (HMI) perspectives. By 2030 we could find greater societal focus on the sustainability, nature of humanity, values, creativity and self/social fulfillment and empowerment of calling for a human-centered computing approach, human centered 6G development. It is essential to consider the wide variety in the humankind and in potential future users of technologies: who they are, what kind of factors help in their inclusion or cause their exclusion, and how we can empower them to critically reflect and evaluate their technology-rich everyday life and role of technology in it, and in the whole society. Unless every person has access to affordable broadband internet and uses it in daily life, the mission of achieving the UN SDGs would remain a distant dream. Therefore, this should be a key requirement as 6G is developed. All those users who still do not use the Internet by choice or for other reasons (e.g., lack of access, affordability, digital literacy) should be offered incentives to experience Internet for free. This would let them know how it can improve their quality of life. Governments and the service providers need to launch public awareness campaigns and make investment e.g. in every school to have digital connectivity.

**Mobile network operators -** The MNOs need to realize that meeting the UN SDGs would increase the size of the total available market size with possibilities to offer completely new breed of applications and services at a price point that would be affordable and yet be profitable because of the increased customer base, increased brand loyalty and potentially increasing the pricing plans as the subscribers become wealthier and better educated. Rather than planning and deploying the same network solution in the rural and under-connected regions as in the urban metropolitan areas, they ought to employ a more flexible approach to network design customized for different rural and suburban section of the population, such as the terrain, availability of power grid, demographics and capacity to absorb the digital technologies. Some examples are the use of shared spectrum, edge computing/caching, renewable energy, and crowd sourcing to build a more cost-effective network architecture. Partnering with the government, industry and local community would ensure that everybody's expectations are met in a sustainable, secured and trustworthy manner. An eco-system of local operators to complement existing MNOs needs to flourish to enable scale and faster time to deployment with equitable revenue sharing models.

**Network equipment manufacturers -** Although there are equipment manufactures of all technologies, they are generally optimized for urban areas and the digitally literate section of the society serving a large number of subscribers. The manufacturers should cost-optimize these network elements for low-density sparse service area rather than strip down the equipment designed for urban metropolitan users. To keep the capital cost of network equipment down, there would probably need to be a better cost sharing between manufacturer and the service provider such that the equipment is paid over a period of time than upfront. There could possibly be innovation in business model as well where revenue is shared between the two.

**Application and service providers –** 6G will enable a new set of applications and services that are based on 6G networks' ability to merge communication service with e.g. accurate position service and high-resolution imaging service. For the various users of 6G in developed and developing worlds, the UIs need to be more user friendly, with audio/video-based interaction, automatic language translation and minimizing text-based (involving read-write) input-output, primarily for those who are not literate. Solution to authentication and security must be simplified to match the digital capacity of the users, especially in the developing world.

There is also a need to move away from charging the user for the application to charging the content provider and/or relying more on advertising revenue. Many of the applications helping the users improve their quality of life (through support of many of the UN SDGs) can be made available for free from the service provider cloud (called digital public goods, DPGs) or the government owned content servers in the cloud.

**Verticals –** Vertical industries and also their public sector counterparts including stakeholders from e.g. automotive, healthcare, energy and other sectors are the future users of the 6G technology. Adopting the national and international level agreements for inherent changes that need to take place to meet the UN SDGs places significant economical constraints on the verticals. To transform operations to meet the UN SDGs, the verticals will need to take all from the future technologies can offer. Therefore, verticals need to engage early on in the process of 6G development and not wait for the telecommunications industry to define what 6G can bring for verticals. Also, special purpose 6G networks will be operated in the verticals and the various applications tailored for verticals' use will be running on the 6G networks. Therefore, the vertical could take an active role in 6G.

**Business Drivers -** There is a need to think of out-of-the-box and create new collaborative business models. For example, these could include users as part of the revenue eco-system, federated deployment and interconnection of user equipment, a barter system of credits in money and goods, direct negotiation between users and content providers or industry verticals. Another good example is emerging in the power industry with micro-grids, which ought to be looked into in the telecom industry. Freemium models and free access to internet for basic content (i.e., non-video content) would help accelerate subscriber base among the digitally disadvantaged. Free internet in areas of extreme poverty would increase the possibility of education and business opportunities for people without means. Use of cloud, block chain, edge computing and AI should be leveraged to bring about a complete transformation in the mobile industry where it is democratized with some control given to the users. At the present time, the industry, i.e., the mobile service providers have a monopolistic control on the subscriber base. This has been broken to some extent with the broadband mobile internet due to net-neutrality. But, still this has a long way to go when it comes to network infrastructure deployment and revenue sharing.

# 7. Concluding remarks

This white paper has launched the process on developing a linkage between 6G and the UN SDGs that are both targeted for 2030. This has turned out to be a complex task as the UN SDG framework comprehensively considers societal challenges and the development of 6G is still at an early stage. However, the role of telecommunications is inevitable in aiming to meet the UN SDGs. To bridge the gap, the role of 6G is here proposed to be seen to consist of three distinct viewpoints: 6G as 1) a provider of services to help steer communities and countries towards reaching the UN SDGs', 2) an enabler of measuring tool for data collection to help reporting of indicators with hyperlocal granularity, and 3) a reinforcer of a new 6G ecosystem to be developed in line with the UN SDGs. By looking into where 6G can impact the indicators defined in the UN SDG framework from these viewpoints, we have outlined connections between 6G and the UN SDGs. This white paper has also started to derive a new set of indicators for 6G which are not yet included in the UN SDG framework or in the development of prior generations of mobile communication networks. A number of research questions for the future remains to be explored including the following:

- What is the role of 6G in meeting the UN SDGs?
- What are proper indicators for 6G?
- How can public and private sector work together to build most powerful 6G vision with maximum UN SDG impact?
- How to avoid global fragmentation of both UN SDG and 6G technology evolution?
- What roles should the different stakeholders take in the sustainable development of 6G?
- What is the role of researchers and the research infrastructures and practices in the development of sustainable future with 6G?
- How to optimize the development of 6G and related technologies in the specific areas of the UN SDGs such as climate change?
- How to achieve the goals related to human and society development and create bridges between human sciences, people as users of the technology and 6G developers in academia and beyond?
- How to empower the young generation to contribute to human-centered 6G development targeting the UN SDGs and the needs of those being marginalized or disadvantaged?